\documentclass[10pt]{article}
\usepackage{CJK}
\usepackage{bbm}
\usepackage{bm}
\usepackage{amsmath}
\usepackage{upgreek}
\usepackage{mathbbold}
\usepackage{amsmath,amssymb,color}
\usepackage{amsfonts}
\usepackage[all]{xy}
\usepackage{graphics}

\newcommand{\bmnabla}{{\boldsymbol{\nabla}}}

\newcommand{\ii}{{\rm i}}
\newcommand{\ep}{{\rm e}}
\newcommand{\be}{\begin{equation}}
\newcommand{\ee}{\end{equation}}
\newcommand{\bea}{\setlength\arraycolsep{2pt} \begin{eqnarray}}
\newcommand{\eea}{\end{eqnarray}}

\newcommand{\rd}{{\rm d}}
\newcommand{\rD}{{\rm D}}

\newcommand{\mPhi}{{\mathbbold{\Phi}}}
\newcommand{\mGamma}{{\mathbbold{\Gamma}}}

\newcommand{\halpha}{{\hat\alpha}}
\newcommand{\hbeta}{{\hat\beta}}

\newcommand{\bu}{{\mathbbm u}}

\newcommand{\mh}{{\mathbbm h}}
\newcommand{\mD}{{\mathbbm D}}
\newcommand{\md}{{\mathbbm d}}

\newcommand{\mY}{{\mathbbm Y}}

\newcommand{\mK}{{\mathbbm K}}

\newcommand{\mJ}{{\mathbbm J}}
\newcommand{\mH}{{\mathbbm H}}
\newcommand{\mR}{{\mathbbold R}}

\newcommand{\wD}{{\mathcal D}}

\newcommand{\mF}{{\mathbbm F}}
\newcommand{\mg}{{\mathbbold{g}}}

\newcommand{\cA}{{\mathcal A}}

\newcommand{\bmA}{{\bm A}}

\newcommand{\bmJ}{{\bm J}}

\newcommand{\bfD}{{\mathbf D}}

\newcommand{\bmR}{{\bm R}}
\newcommand{\bmY}{{\bm Y}}
\newcommand{\bmV}{{\bm V}}
\newcommand{\fn}{{\mathfrak n}}

\newcommand{\bmT}{{\bm T}}

\newcommand{\bmg}{{\bm g}}
\newcommand{\bmt}{{\bm t}}

\newcommand{\bmF}{{\bm F}}
\newcommand{\hmf}{{\bm f}}

\newcommand{\bmX}{{\bm X}}

\newcommand{\mA}{{\mathbbm A}}

\newcommand{\mN}{{\mathbbm N}}

\newcommand{\mt}{{\mathbbm t}}

\newcommand{\cD}{{\mathcal D}}

\newcommand{\bcD}{{\mathbbold{D}}}

\renewcommand{\thefootnote}{\fnsymbol{footnote}}

\begin{document}

\makeatletter
\@addtoreset{equation}{section}
\makeatother
\renewcommand{\theequation}{\arabic{section}.\arabic{equation}}
\begin{titlepage}
\vfill
\begin{flushright}

\end{flushright}
\vfill
\begin{center}
{\Large\bf AdS gravity, $SO(2,d)$ gauge theory \\ and Holography}

\vskip 1cm

Zhao-Long Wang$^{1,2,3}$\footnote{\tt zlwang@nwu.edu.cn},
Ya-Xun Song$^{1}$

\vskip 5mm
$^1${\it Institute of Modern Physics, Northwest University, XiAn 710127, China}
\\$^2${\it Peng Huanwu Center for Fundamental Theory, XiAn 710127, China}
\\$^3${\it Shaanxi Key Laboratory for Theoretical Physics Frontiers, XiAn 710127, China}

\end{center}
\vfill

\begin{abstract}
\noindent
Inspired by the general relation between the boundary global symmetry and the bulk gauge symmetry in AdS/CFT, we reformulate the $d+1$ dimensional AdS gravity theory as a $SO(2,d)$ gauge theory. In this formalism, the pull back of the bulk equation of motion onto a co-dimension one hypersurface $\Sigma$ can be naturally related to the $SO(2,d)$ conservation law under a local energy scale in the dual CFT.
The consistency of CFT requires that the $SO(2,d)$ covariant anomaly vanishes at arbitrary scale. After imposing a covariant area law as the renormalization condition, the vanishing of $SO(2,d)$ anomaly implies that the dual bulk geometries must satisfy the Einstein equation.
\end{abstract}

\vfill
\end{titlepage}

\tableofcontents\newpage
\renewcommand{\thefootnote}{\#\arabic{footnote}}
\setcounter{footnote}{0}

\section{Introduction}
The AdS/CFT correspondence \cite{Maldacena:1997re,Gubser:1998bc,Witten:1998qj} implies a duality between the quantum gravity in $D=d+1$ dimensional anti-de Sitter space and the $d$-dimensional conformal field theory. The holographic dictionary between the boundary data of asymptotic AdS space and the CFT quantities has been well established in \cite{Witten:1998qj} by the field-operator correspondence.
However, apart form the AdS boundary, the explicit CFT realization of the bulk local degree of freedom remains unclear yet. Especially, there is no systematical explanation on the emergence of bulk Einstein equation from the CFT side. Various approaches have been proposed on the related topics in the past twenty years, including the holographic Callan-Symanzik equation \cite{de Boer:1999xf}, the holographic Wilsonian  renormalization group \cite{Heemskerk:2010hk},  the smearing operator \cite{Kabat:2011rz}, the tensor networks \cite{Swingle:2009bg}, the integral geometry \cite{Czech:2015qta}, etc. In our early work \cite{Wang:2015qfa}, by considering the conformal transformation of the renormalization scale, it was shown that the bulk dynamics of a scalar field is highly constrained by the $SO(2,d)$ conformal symmetry of the dual CFT scalar operators. If we consider the CFT energy momentum tensor instead, it is natural to explore whether the corresponding bulk gravity dynamics is also constrained by the $SO(2,d)$ symmetry.

Different from the simple scalar operator considered in \cite{Wang:2015qfa},  the energy momentum tensor itself is related to the conformal Noether current. In such kind of cases, a general principle was noticed in the studies of the AdS/CFT. That is, the global symmetry in the boundary field theory is dual to the gauge symmetry in the bulk theory. It plays a very important role in the applications of AdS/CFT, e.g., the holographic superconductor \cite{Hartnoll:2008vx}. Applying this principle to the conformal symmetry in the boundary CFT, we would expect the dual bulk theory is a gauge theory with the gauge group $SO(2,d)$. However, only the $SO(1,d)$ local Lorentzian gauge symmetry is manifest  in the usual formula of bulk gravity. To understand the relation between the $SO(2,d)$ symmetry and the bulk dynamics,  we need to find a uplifted action of gravity in which the $SO(2,d)$ local gauge symmetry appears manifestly. For $D=3$, it is already well known that the AdS$_3$ gravity can be reformulated as a $SO(2,2)=SL(2,R)\times SL(2,R)$ Chern-Simons theory \cite{Witten:1988hc}. This paper is started from showing the similar uplift can be established for general dimensions in Section 2.1. In Section 2.2, the relation to the traditional $SO(2,d)$ invariant expressions is analysed by introducing an intrinsic $SO(2,d)$ basis.  Then in Section 3.1, the corresponding Hamiltonian formalism  is analysed. In this formalism, the pull back of the bulk equation of motion onto a co-dimension one hypersurface $\Sigma$ can be naturally related to the anomaly of $SO(2,d)$ conservation law. Providing the $SO(2,d)$ conservation law is not anomalous on arbitrary $\Sigma$, all components of bulk Einstein equation will be automatically satisfied.  In Section 3.2.1 and 3.2.2, we sketch how to establish the corresponding generic $SO(2,d)$ background fields description for the CFT. Based on this description, the bulk canonical structure and Hamiltonian constraints are naturally realized in the CFT language in 3.2.3. As result, the bulk dynamic is naturally emergent from the CFT $SO(2,d)$ non-anomalous condition. In section 3.3, we discuss the validity of our construction in more generic bulk theories. In Section 4, we summarize our results and discuss possible generalizations.

\section{AdS gravity as $SO(2,d)$ gauge theory}
\subsection{$SO(2,d)$ uplifting of Einstein gravity}
Written in terms of the vielbein formalism, the $D$-dimensional Einstein-Hilbert action with negative cosmological constant is given by
\begin{eqnarray}\label{2ndGR}
S[e^{a}]\!\!\!\!&=&\!\!\!\!\frac{1}{2\kappa^2\,(D-2)!}\int\epsilon_{a_1\cdots a_D} \left[\Theta^{a_1a_2}+\frac{(D-2)}{D\,\ell^2} e^{a_1}\wedge e^{a_2}\right]\wedge e^{a_3}\wedge\cdots \wedge e^{a_D}\,,
\end{eqnarray}
where $\epsilon_{a_1\cdots a_D}$ is the unit total antisymmetric tensor of $SO(1,d)$ vielbein indices $a_i=0,\cdots,d$. In the usual understanding of Einstein gravity, the fundamental dynamical field is just the vielbein 1-form $e^a=e^{a}{}_{M}\rd x^{M}$ where $\{x^{M}\}=\{x^{\mu},z\}$ are the bulk spacetime coordinates.
The curvature 2-form
\begin{eqnarray}
\Theta^{a}{}_{b}\!\!\!\!&=&\!\!\!\!\rd\omega^{a}{}_{b}+\omega^{a}{}_{c}\wedge\omega^{c}{}_{b}=\tfrac12R^{a}{}_{bMN}\rd x^{M}\wedge\rd x^{N}\,,
\end{eqnarray}
is the field strength of the spin connection 1-form $\omega^{ab}=-\omega^{ba}$ which is just the gauge field of the local $SO(1,d)$ group. Given the torsion free condition
\begin{eqnarray}\label{TF2}
\rD e^{a}=\rd e^a+\omega^{a}{}_{b}\wedge e^b=0\,,
\end{eqnarray}
the spin connection is totally fixed by the vielbein.
Taking the variation of the action (\ref{2ndGR}), we get the Einstein equation
\begin{eqnarray}
\left(\Theta^{[a_1a_2}+\ell^{-2}e^{[a_1}\wedge e^{a_2}\right)
\wedge e^{a_3}\wedge\cdots\wedge e^{a_{D-1}]}=0\,.
\end{eqnarray}

Alternatively, in the Palatini understanding of gravity, both the vielbein and the spin connection are regarded  as independent variables
\begin{eqnarray}
S[e^{a},\omega^{a}{}_{b}]\!\!\!\!&=&\!\!\!\!\frac{1}{2\kappa^2\,(D-2)!}\int\epsilon_{a_1\cdots a_D} \left[\Theta^{a_1a_2}+\frac{(D-2)}{D\,\ell^2} e^{a_1}\wedge e^{a_2}\right]\wedge e^{a_3}\wedge\cdots \wedge e^{a_D}\,.~~~
\end{eqnarray}
The corresponding EOM's from the variation of $e^a$ and $\omega^{ab}$ are respectively
\begin{eqnarray}\label{Ein1}
\left(\Theta^{[a_1a_2}+\ell^{-2}e^{[a_1}\wedge e^{a_2}\right)
\wedge e^{a_3}\wedge\cdots\wedge e^{a_{D-1}]}=0\,,
\\\label{TF1}
\rD e^{[a_1}\wedge e^{a_{2}}\wedge\cdots \wedge e^{a_{D-2}]}=0\,.
\end{eqnarray}
Providing that the vielbein $e^{a}$ is not degenerate, the second equation (\ref{TF1}) is equivalent to the torsion free condition (\ref{TF2}). Therefore, the Palatini action is equivalent to the original AdS gravity classically.

How to reformulate the above theory as a $SO(2,d)$ gauge theory?
A natural idea is to regard the vielbein $e^a$ and the spin connection $\omega^{ab}$ as different components of the $SO(2,d)$ gauge field ${A}^{\hat\alpha\hat\beta}$. That is
\begin{eqnarray}\label{initi}
{A}^{\hat a\hat b}=\omega^{ab}\,,~~~{A}^{\hat a\hat\bullet}=\ell^{-1} e^{a}=-{A}^{\hat\bullet\hat a}\,,
\end{eqnarray}
where we use $\hat\bullet$ to denote the additional indices of $SO(2,d)$ vector space and $\hat\alpha,\hat\beta$ are the vector indices of $SO(2,d)$ group.
Then the corresponding field strength is related to the curvature as well as the torsion
\begin{eqnarray}
{F}^{ab}\!\!\!\!&=&\!\!\!\!
\rd{A}^{ab}+{A}^{a}{}_{\hat\gamma}\wedge{A}^{\hat\gamma b}
=\rd\omega^{ab}+\omega^{a}{}_c\wedge\omega^{cb}+\ell^{-2}e^{a}\wedge e^{b}
=\Theta^{ab}+\ell^{-2}e^{a}\wedge e^{b}\,,~~~
\cr
{F}^{a\hat\bullet}\!\!\!\!&=&\!\!\!\!
\rd{A}^{a\hat\bullet}+{A}^{a}{}_{\hat\gamma}\wedge{A}^{\hat\gamma\hat\bullet}
=\ell^{-1}(\rd e^a+\omega^{a}{}_c\wedge e^{c})=\ell^{-1}\,\wD e^a\,.
\end{eqnarray}
More systematically, to split the $SO(2,d)$ gauge connection to $e^a$ and $\omega^{ab}$, we need to introduce an additional field $Y^{\hat\alpha}$ which is in the vector representation of $SO(2,d)$. Furthermore, we can impose the following gauge invariant constraint on the $Y^{\hat\alpha}$ field
\begin{eqnarray}
Y^{\hat\alpha}Y_{\hat\alpha}=-\ell^2\,,
\end{eqnarray}
such that it can be totally fixed by the $SO(2,d)$ gauge choice and does not introduce any additional physical degree of freedom.
In this general set up, the space-time metric is given by the following gauge invariant quadratic form
\begin{eqnarray}
g_{MN}\!\!\!\!&=&\!\!\!\!\rD_{M}Y^{\hat\alpha}\rD_{N}Y_{\hat\alpha}\,,
\end{eqnarray}
where $\rD=\rd+A$ denotes the $SO(2,d)$ gauge covariant derivative. The $Y^{\hat\alpha}$ field just tells us how to induce the spacetime metric out of the inner $SO(2,d)$ gauge field configuration. Thus we shall refer the $Y^{\hat\alpha}$ as the ``ruler field''.

The initial idea of $SO(2,d)$ uplifting (\ref{initi}) can be regarded as the special formula under the gauge choice(the Einstein gauge)
\begin{eqnarray}\label{Egauge}
Y_{\hat a}=0\,,~~~~Y_{\hat\bullet}=\ell\,.
\end{eqnarray}
In this special gauge, we have
\begin{eqnarray}
\rD Y^{\hat a}\!\!\!\!&=&\!\!\!\!\rd Y^{\hat a}+{A}^{\hat a}{}_{\hat\beta}Y^{\hat\beta}=\ell{A}^{\hat a\hat\bullet}=e^a\,,~~~~~~
\rD Y^{\hat\bullet}=\rd Y^{\hat\bullet}+{A}^{\hat\bullet}{}_{\hat\beta}Y^{\hat\beta}=0\,,
\cr \rD\rD Y^{\hat a}\!\!\!\!&=&\!\!\!\!{F}^{\hat a}{}_{\hat\beta}Y^{\hat\beta}
=\ell{F}^{\hat a\hat\bullet}=\wD e^a\,,~~~~~~~~~~
\rD\rD Y^{\hat\bullet}={F}^{\hat\bullet}{}_{\hat b}Y^{\hat b}=0\,.
\end{eqnarray}
Now the Palatini EOMs (\ref{Ein1}) and (\ref{TF1})
can be nicely unified in a $SO(2,d)$ covariant way
\begin{eqnarray}
 F^{[\hat\alpha_1\hat\alpha_2}\wedge\rD Y^{\hat\alpha_3}\wedge\cdots\wedge \rD Y^{\hat\alpha_{d}]}=0\,.
\end{eqnarray}

One can further realize the uplifting at the action level. In the Einstein gauge (\ref{Egauge}), we have
\begin{eqnarray}
\!\!\!\!&&\!\!\!\!\left(\Theta^{[a_1a_2}+\ell^{-2}e^{[a_1}\wedge e^{a_2}\right)
\wedge e^{a_3}\wedge\cdots\wedge e^{a_{D}]}
\cr \!\!\!\!&=&\!\!\!\!(D+1)\,\ell^{-1} F^{[\hat\alpha_1\hat\alpha_2}\wedge \rD Y^{\hat\alpha_3}\wedge\cdots\wedge \rD Y^{\hat\alpha_{D}}Y^{\hat\alpha_{D+1}]}\,,
\cr\!\!\!\!&&\!\!\!\!e^{[a_1}\wedge e^{a_2}
\wedge e^{a_3}\wedge\cdots\wedge e^{a_{D}]}
\cr \!\!\!\!&=&\!\!\!\!(D+1)\ell^{-1}\rD Y^{[\hat\alpha_1}\wedge \rD Y^{\hat\alpha_2}\wedge\cdots\wedge \rD Y^{\hat\alpha_{D}}Y^{\hat\alpha_{D+1}]}\,.
\end{eqnarray}
It suggests the following gauge invariant action
\begin{eqnarray}\label{GRGT}
S[Y,A]
\!\!\!\!&=&\!\!\!\!\frac{1}{2\kappa^2\ell\,(D-2)!}\int_{M} \epsilon_{\hat\alpha_1\cdots \hat\alpha_{D+1}} \left[ F^{\hat\alpha_1\hat\alpha_2}-\frac{2}{D\ell^2} \rD Y^{\hat\alpha_1}\wedge  \rD Y^{\hat\alpha_2}\right]\wedge  \rD Y^{\hat\alpha_3}\wedge\cdots \wedge  \rD Y^{\hat\alpha_D}Y^{\hat\alpha_{D+1}}
\,.~~~~~~~~~~
\end{eqnarray}
For $D=4${\footnote{After this paper was completed, we noticed that the $D=4$ version of the action (\ref{GRGT}) has already appeared in Ref.\cite{Wilczek:1998ea}.}}, the difference between (\ref{GRGT}) and the MacDowell-Mansouri-Stelle-West action\cite{MacDowell:1977jt,Mansouri:1977ej,Stelle:1979aj} is just the Gauss-Bonnet term. Since the 4d Gauss-Bonnet is topological, these two actions are equivalent to each other and give rises to same equations of motion.  Does it also equivalent to the Chern-Simons action \cite{Witten:1988hc} in $D=3$?
At the first sight, the above action looks quite different from the Chern-Simons type of action. Especially, the $Y^{\hat\alpha}$ field does not appear in \cite{Witten:1988hc}.  However, this is just an illusion  due to the fact that $Y^{\hat\alpha}$ does not appears in the EOM in $D=3$. In fact, the $Y^{\hat\alpha}$ is implicity imposed in \cite{Witten:1988hc} when one try to identify the vielbein $e^{a}$ with a specific linear combination $A_L+A_R$ of the $SL(2,R)\times SL(2,R)$ gauge field. Furthermore,
the equivalence can be proved by noticing that these two  actions are differed by a total derivative term.
In general, one can write down the higher dimensional analogy of the action in \cite{Witten:1988hc} and prove that these two types of actions are always equivalent to each other up to a total derivative term. The details are explained in the Appendix A.

By varying the gauge field $A^{\hat\alpha\hat\beta}$ in (\ref{GRGT}),  we get
\begin{eqnarray}\label{EOM1}
(D-2)\,F^{[\hat\alpha_1\hat\alpha_{2}}\wedge\rD Y^{\hat\alpha_3}\wedge\cdots\wedge\rD Y^{\hat\alpha_{D-1}]}=0\,.~~~~~~~~~
\end{eqnarray}
It explicitly reproduce the expected EOM for $D>2$.
For $D=2$, the EOM is trivial since (\ref{GRGT}) becomes a total derivative term. 
On the other hand, the EOM derived from varying $Y^{\hat\alpha}$ field{\footnote{Since the norm of $Y^{\hat\alpha}$ is fixed, the independent components of $\delta Y$ EOMs are given by $(\delta_{\hat\alpha}^{\hat\beta}+\ell^2 Y_{\hat\alpha}Y^{\hat\beta})\frac{\delta S}{\delta Y^{\hat\beta}}$ which are orthogonal to $Y^{\hat\alpha}$.}} is
\begin{eqnarray}\label{EOM2}
(D-2)\,\rD\!\left(F^{[\hat\alpha_1\hat\alpha_{2}}\wedge\rD Y^{\hat\alpha_3}\wedge\cdots\wedge\rD Y^{\hat\alpha_{D-1}}\right)Y^{\hat\alpha_D]}=0\,.
\end{eqnarray}
When (\ref{EOM1}) is satisfied, (\ref{EOM2}) is automatically satisfied. Thus as expected, the introducing of $Y^{\hat\alpha}$ does not imposing any additional constraints other than the original Palatini equations.

Besides the Einstein gauge, another useful gauge choice is
\begin{eqnarray}\label{bulkEmG}
Y^{\hat\mu}(x,z)\!\!\!\!&=&\!\!\!\! \frac{x^\mu}{ z} \,, ~~~
Y^{\hat d}(x,z)=  \frac{ \ell^2-\eta_{\mu\nu}x^\mu x^\nu-\ell^2 z^2}{2\ell z}\, ,
~~~ Y^{\hat\bullet}(x,z)=\frac{\ell^2+\eta_{\mu\nu}x^\mu x^\nu+\ell^2 z^2}{ 2\ell z}\,,~~~~
\end{eqnarray}
where $\mu=0,\cdots,d-1$. We will denote it as the embedding gauge since $Y^{\hat\alpha}$ takes the value of embedding coordinates of pure AdS in $D+1$ dimensional flat space.
In this gauge, the $A=0$ configuration gives rise to the pure AdS vacuum
\begin{eqnarray}\label{ads}
  {\rd s}^2 = \frac{1}{z^2}\left({\ell^2{\rd z}^2}    + \eta_{\mu\nu}{\rd x}^{\mu}{\rd x}^{\nu}\right)\,.
\end{eqnarray}
Fixing in the embedding gauge (\ref{bulkEmG}), a coordinate transformation on $Y^{\hat\alpha}$ can be mapped to a $SO(2,d)/SO(1,d)$ gauge transformation which is decided up to the $SO(1,d)$ subgroup leaving $Y^{\hat\alpha}$ intact. Especially, the isometries of pure AdS vacuum are mapped to the rigid $SO(2,d)$ transformations.
\subsection{The intrinsic $SO(2,d)$ basis}
To clarify the relation between the above gauge theory notations and the usual $SO(2,d)$ invariant notations of gravity, let us expand the quantities with $SO(2,d)$ indices by a gauge covariant basis. We notice that
\begin{eqnarray}
Y^{\hat\alpha}Y_{\hat\alpha}\!\!\!\!&=&\!\!\!\!-\ell^2\,,~~~~\rD_{M}Y^{\hat\alpha}\rD_{N}Y_{\hat\alpha}=g_{MN}
\,,~~~~
 Y^{\hat\alpha}\rD_{M } Y_{\hat\alpha}=0\,.~~~
\end{eqnarray}
Thus $\{Y^{\hat\alpha},\rD_{M }Y^{\hat\alpha}\}$ naturally forms a orthogonal basis of the $SO(2,d)$ vector space when $\rD Y^{\hat\alpha}$ is not degenerate.  The completion relation  is obviously
\begin{eqnarray}
\rD_{M}Y^{\hat\alpha}\rD^{M}Y^{\hat\beta}-\ell^{-2} Y^{\hat\alpha} Y^{\hat\beta}=\eta^{\hat\alpha\hat\beta}\,.~~~
\end{eqnarray}
Correspondingly, the intrinsic basis for the adjoint representation is
\begin{eqnarray}
(\tau_{M})^{\hat\alpha}{}_{\hat\beta}\!\!\!\!&=&\!\!\!\!\ell^{-1}\left(Y^{\hat\alpha}\rD_{M} Y_{\hat\beta}- Y_{\hat\beta} \rD_{M}Y^{\hat\alpha}\right)\,,
\cr
(\tau_{MN})^{\hat\alpha}{}_{\hat\beta}\!\!\!\!&=&\!\!\!\!-(\tau_{NM})^{\hat\alpha}{}_{\hat\beta}
=\rD_{M}Y^{\hat\alpha} \rD_{N} Y_{\hat\beta}-\rD_{M} Y_{\hat\beta} \rD_{N}Y^{\hat\alpha}
\,.
\end{eqnarray}
The commutators
\begin{eqnarray}
\!\!\!\!&&\!\!\!\![\tau_{M},\tau_{N}]=-\tau_{MN}\,,~~~~~~~~~
[\tau_{M_1M_2},\tau_{N}]=-2g_{N[M_1}\tau_{M_2]}\,,
\cr\!\!\!\!&&\!\!\!\!
[\tau_{M_1M_2},\tau_{N_1N_2}]
=2\left(g_{N_1[M_2}\tau_{M_1]N_2}-g_{N_2[M_2}\tau_{M_1]N_1}\right)  \,,
\end{eqnarray}
give rise to a local realization of the $so(2,d)$ Lie-algebra since it dependents on the local metric $g_{MN}$ manifestly .

Now we can expand the $SO(2,d)$ gauge theory quantities in the intrinsic basis.  We notice that
\begin{eqnarray}
 Y^{\hat\alpha}\rD_{N}\rD_{M } Y_{\hat\alpha}\!\!\!\!&=&\!\!\!\!-g_{M N}\,,
\cr
\rD_{N}Y^{\hat\alpha}\rD_{(M_2}\rD_{M_1)} Y_{\hat\alpha}
\!\!\!\!&=&\!\!\!\!\frac12\left[\rD_{M_2}(\rD_{N}Y^{\hat\alpha}\rD_{M_1} Y_{\hat\alpha})+\rD_{M_1}(\rD_{N}Y^{\hat\alpha}\rD_{M_2} Y_{\hat\alpha})
-\rD_{N}(\rD_{M_1}Y^{\hat\alpha}\rD_{M_2} Y_{\hat\alpha})\right]
\cr\!\!\!\!&&\!\!\!\!+\rD_{[N}\rD_{M_1]}Y^{\hat\alpha}\rD_{M_2} Y_{\hat\alpha}
+\rD_{M_1}Y^{\hat\alpha}\rD_{[N}\rD_{M_2]} Y_{\hat\alpha}
\cr\!\!\!\!&=&\!\!\!\!\frac12\left(\partial_{M_2}g_{N M_1}+\partial_{M_1}g_{NM_2}
-\partial_{N}g_{M_1M_2}\right)+(F_{N(M_1})^{\hat\alpha\hat\beta}\rD_{M_2)}Y_{\hat\alpha}\,Y_{\hat\beta}
\,,~~~
\cr
\rD_{N}Y^{\hat\alpha}\rD_{[M_2}\rD_{M_1]} Y_{\hat\alpha}
\!\!\!\!&=&\!\!\!\!\frac12(F_{M_2M_1})^{\hat\alpha\hat\beta}Y_{\hat\beta}\rD_{N}Y_{\hat\alpha}\,.
\end{eqnarray}
Thus we have the expansion of $\rD\rD Y$ as
\begin{eqnarray}\label{bulkdev}
\rD_{M_2}\rD_{M_1} Y^{\hat\alpha}
\!\!\!\!&=&\!\!\!\!\ell^{-2} g_{M_1 M_2}Y^{\hat\alpha}+\Gamma^{N}{}_{M_1M_2}\rD_{N}Y^{\hat\alpha}\,,~~~~~~~~
\end{eqnarray}
which suggests the following spacetime connection $\Gamma^{M}{}_{NP}$ with torsion $t^{M}{}_{NP}$
\begin{eqnarray}
\Gamma^{M}{}_{NP}\!\!\!\!&=&\!\!\!\!\hat\Gamma^{M}{}_{NP}+{t}^M{}_{NP}\,,
\cr
\hat\Gamma_{MM_1M_2}\!\!\!\!&=&\!\!\!\!\frac12\left(\partial_{M_2}g_{N M_1}+\partial_{M_1}g_{NM_2}-\partial_{N}g_{M_1M_2}\right)\,,
\cr
{t}_{NM_1M_2}
\!\!\!\!&=&\!\!\!\!\frac{1}{2}\left[(F_{NM_1})^{\hat\beta_1\hat\beta_2}\rD_{M_2}Y_{\hat\beta_1}
+(F_{NM_2})^{\hat\beta_1\hat\beta_2}\rD_{M_1}Y_{\hat\beta_1} -(F_{M_1M_2})^{\hat\beta_1\hat\beta_2}\rD_{N}Y_{\hat\beta_1}
\right]Y_{\hat\beta_2}
\,.~~~~~~~~
\end{eqnarray}
Now the torsion free condition can be expressed covariantly as
\begin{eqnarray}
\rD_{[M_1}\rD_{M_2]} Y^{\hat\alpha}
\!\!\!\!&=&\!\!\!\!\frac12(F_{M_1 M_2}){}^{\hat\alpha}{}_{\hat\beta}Y^{\hat\beta}=0\,.~~~~~~~~
\end{eqnarray}
Using the corresponding covariant derivative $\cD=\partial+\Gamma+A=\nabla+A$, we get
\begin{eqnarray}
\cD_{M_2}\cD_{M_1} Y^{\hat\alpha}
\!\!\!\!&=&\!\!\!\!\ell^{-2} g_{M_1 M_2}Y^{\hat\alpha}\,,~~~~~~~~
\cr
\cD_{M}\tau_{N}\!\!\!\!&=&\!\!\!\!\ell^{-1}\,\tau_{MN}\,,
\cr
\cD_{M}\tau_{N_1N_2}\!\!\!\!&=&\!\!\!\! 2\ell^{-1}\, g_{M[N_1}\tau_{N_2]}
\,.
\end{eqnarray}

Similarly, by computing the gauge invariant quantities like $\cD_{M_1}Y^{\hat\alpha}\cD_{[M_2}\cD_{M_3]}\cD_{M_4}Y_{\hat\alpha}$, we find the $SO(2,d)$ field strength can be expanded as
\begin{eqnarray}\label{flux}
\!\!\!\!&&\!\!\!\!F_{M_1M_2}
\cr
\!\!\!\!&=&\!\!\!\!\left(\frac12R^{N_1N_2}{}_{M_1M_2} +\ell^{-2}\delta^{N_1}_{[M_1}\delta^{N_2}_{M_2]}\right)\tau_{N_1N_2}-2\ell^{-1}t^{N}{}_{[M_1M_2]}\tau_{N}
\cr\!\!\!\!&=&\!\!\!\!\left(\frac12\hat R^{N_1N_2}{}_{M_1M_2}+\hat\nabla_{[M_1}t^{N_1N_2}{}_{M_2]}
+t^{N_1}{}_{N[M_1}t^{N N_2}{}_{ M_2]} +\ell^{-2}\delta^{N_1}_{[M_1}\delta^{N_2}_{M_2]}\right)\tau_{N_1N_2}-2\ell^{-1}t^{N}{}_{[M_1M_2]}\tau_{N}\,,~~~~~~~~
\end{eqnarray}
where $\hat\nabla$ is the covariant derivative without torsion and $\hat R^{N_1N_2}{}_{M_1M_2}$ is the corresponding curvature tensor.  Expanding the Bianchi identity on the intrinsic basis, we get
\begin{eqnarray}
0\!\!\!\!&=&\!\!\!\!\rD_{[M_3}F_{M_1M_2]}
\cr\!\!\!\!&=&\!\!\!\!
\left(\hat\nabla_{[M_3} \hat R^{N_1N_2}{}_{M_1M_2]}-\hat R^N{}_{[M_2M_3M_1]} t^{N_1N_2}{}_{N} \right)\rD_{N_1}Y^{[\hat\alpha}\rD_{N_2}Y^{\hat\beta]}
-2\ell^{-2}\hat R^{N}{}_{[M_3M_1M_2]}Y^{[\hat\alpha}\rD_{N}Y^{\hat\beta]}\,.~~~~~~~~~
\end{eqnarray}
It is equivalent to the two Bianchi identities for the usual Riemann curvature.

We also notice that
\begin{eqnarray}\label{ddual}
\epsilon_{\hat\alpha_1\cdots\hat\alpha_{D+1}}
\rD_{M_1} Y^{\hat\alpha_1}\cdots \rD_{M_D} Y^{\hat\alpha_{D}} Y^{\hat\alpha_{D+1}} \!\!\!\!&=&\!\!\!\! \ell g^{\frac12}\varepsilon_{M_1\cdots M_D}
\end{eqnarray}
where
\begin{eqnarray}
\varepsilon_{01\cdots d}=-1\,,~~~~g=-\det(g_{MN})\,.
\end{eqnarray}
By using (\ref{ddual}) as well as the expansion (\ref{flux}),
the bulk EOM
\begin{eqnarray}\label{EOMep}
\epsilon_{\hat\alpha\hat\beta\hat\alpha_1\cdots\hat\alpha_{D-1}}F^{\hat\alpha_1\hat\alpha_{2}}\wedge\rD Y^{\hat\alpha_3}\wedge\cdots\wedge\rD Y^{\hat\alpha_{D-1}}=0 ~~~~~~~~~
\end{eqnarray}
can also be decomposed in terms of the intrinsic basis.
We find that the $\tau_{MN}$ components of (\ref{EOMep}) give rise to the torsion free conditions, and the $\tau_{M}$ components of (\ref{EOMep}) give rise to the usual Einstein equations.

\section{Bulk dynamics from CFT conservation laws}
Given a co-dimension one hypersurface
\begin{eqnarray}
\Sigma=\{(x^{\mu},z)|z=\zeta(x^{\mu})\}
\end{eqnarray}
the pull back of the bulk EOM (\ref{EOMep}) on $\Sigma$ given rise to the $\frac{(d+2)(d+1)}{2}$ constraint equations for the field configuration on this hypersurface, while the other $\frac{(d+2)(d+1)d}{2}$ components of the bulk EOM can be viewed as the evolution equations corresponding to the change of the hypersurface. The LHS of (\ref{EOMep}) is given by a $d-$form, thus the constraint equations is equivalent to
\begin{eqnarray}\label{ProjEin}
 F^{[\hat\alpha_1\hat\alpha_2}\wedge\rD Y^{\hat\alpha_3}\wedge\cdots\wedge \rD Y^{\hat\alpha_{D-1}]}\wedge \mathcal N(\Sigma)=0
\end{eqnarray}
where $\mathcal N(\Sigma)=\rd z-\partial_{\mu}\zeta \rd x^{\mu}$ is the normal 1-form of the hypersurface $\Sigma$. An important observation is
that the full bulk Einstein equations will be automatically satisfied if the constraints (\ref{ProjEin}) are valid on any arbitrary hypersurface $\Sigma$.

It is known that the radial coordinate $\upmu=z^{-1}$ can be explained as the energy scale  in the dual CFT \cite{Maldacena:1997re,Gubser:1998bc,Witten:1998qj}.
Thus given a bulk hypersurface $\Sigma$ is related to define the CFT with a finite and position dependent local energy scale. It should not be surprised to introduce the position dependent local energy scale in CFT.   In fact, according to the general conformal transformation including the energy scale\cite{Wang:2015qfa}, a constant energy scale will become position dependent after performing the special conformal transformation. Furthermore, to compare with the bulk $SO(2,d)$ gauge theory description, one need to introduce the $SO(2,d)$ background field on the CFT side as well. 

In the CFT perspective, the general conservation law for the conformal symmetry also contains $\frac{(d+2)(d+1)}{2}$ components due to the $SO(2,d)$ group structure. As a defining property of the conformal field theory, the $SO(2,d)$ conservation law should be preserved after the renormalization under a local energy scale. Thus it is straightforward to conjecture that these conservation laws are the CFT dual of the constraints (\ref{ProjEin}) from the bulk side. If this conjecture is correct, it means that once we establish the $SO(2,d)$ conservation law for any local energy scale at the CFT side, the dual bulk EOM will be automatically implied. 
\subsection{Bulk Hamiltonian analysis}
\subsubsection{The induced $SO(2,d)$ structure on $\Sigma$}
To explore the relation between the constraint equations on $\Sigma$ and the $SO(2,d)$ conservation law, a natural tool is the Hamiltonian formalism. Let us start the discussions from defining the $d+1$ decomposition of the bulk fields for $\Sigma$.

The projection of the $Y$ field and bulk gauge fields on $\Sigma$ are given by
\begin{eqnarray}
\mY(x;\Sigma)\!\!\!\!&=&\!\!\!\!Y(x,\zeta(x))\,,
\cr\mA_{\mu}(x;\Sigma)\!\!\!\!&=&\!\!\!\!\mh_{\mu}{}^{M}A_M(x,\zeta(x))
=A_{\mu}+\partial_{\mu}\zeta A_{z}\,,~~~~~\mPhi(x;\Sigma)=A_{z}(x,\zeta(x))\,,
\end{eqnarray}
where $\mh_{\mu}{}^{M}$ is the pull back matrix
\begin{eqnarray}
\mh_{\mu}{}^{\nu}=\delta_{\mu}^{\nu}\,,~~~~~\mh_{\mu}{}^{z}=\partial_{\mu}\zeta\,.
\end{eqnarray}
The pull back of the flux is
\begin{eqnarray}
\mF_{\mu\nu}\!\!\!\!&=&\!\!\!\!\mh_{\mu}{}^M\mh_{\nu}{}^NF_{MN}=2\md_{[\mu}\mA_{\nu]}+[\mA_{\mu},\mA_{\nu}]
\,,
\end{eqnarray}
where
\begin{eqnarray}
\md_{\mu}=\mh_{\mu}{}^{M}\partial_{M}
\end{eqnarray}
is the projected derivative on $\Sigma$.
Inversely, we have
\begin{eqnarray}
A=A_{\mu}\rd x^{\mu}+A_{z}\rd z=(\mA_{\mu}-\partial_{\mu}\zeta\mPhi)\rd x^{\mu}+\mPhi\rd z\,,
\end{eqnarray}
and
\begin{eqnarray}
\rD Y^{\hat\alpha}\!\!\!\!&=&\!\!\!\!
(\mD_{\mu} \mY^{\hat\alpha}-\partial_{\mu}\zeta\,\rD_{z} \mY^{\hat\alpha}) \rd x^{\mu}+\rD_{z} \mY^{\hat\alpha}\rd z\,,
\cr F\!\!\!\!&=&\!\!\!\!
(\tfrac12\mF_{\mu\nu}-\partial_{[\mu}\zeta\,\partial_{|z|}\mA_{\nu]}+\partial_{[\mu}\zeta\,\mD_{\nu]}\mPhi)\rd x^{\mu}\wedge\rd x^{\nu}+(\partial_{z}\mA_{\nu}-\mD_{\nu}\mPhi)\rd z\wedge\rd x^{\nu}
\,.
\end{eqnarray}
The pull back of the bulk Einstein equation is simply
\begin{eqnarray}\label{mProjEin}
(\mF_{\mu_1\mu_2})^{[\hat\alpha_1\hat\alpha_2}\mD_{\mu_3} \mY^{\hat\alpha_3}\cdots\mD_{\mu_d} \mY^{\hat\alpha_d]}=0\,.
\end{eqnarray}

Just like the bulk metric, the induced metric on $\Sigma$ can be identified as the following $SO(2,d)$ invariant quadratic form
\begin{eqnarray}
\mh_{\mu\nu}=\mh_{\mu}{}^M\mh_{\nu}{}^Ng_{MN}=\mD_{\mu}\mY^{\hat\alpha}\mD_{\nu}\mY_{\hat\alpha}\,.
\end{eqnarray}
One can also try to formulate the intrinsic $SO(2,d)$ basis by $\mY^{\hat\alpha}$ and its derivatives. In additional to $\mY^{\hat\alpha}$ and $\mD_{\mu}\mY^{\hat\alpha}$, we still need to introduce a $\mN^{\hat\alpha}$ satisfying
\begin{eqnarray}\label{mN}
\mN^{\hat\alpha}\mY_{\hat\alpha}=0\,,~~~~~\mN^{\hat\alpha}\mD_{\mu}\mY_{\hat\alpha}=0\,,~~~~~ \mN^{\hat\alpha}\mN_{\hat\alpha}=\ell^2\,.
\end{eqnarray}
Providing $\mD_{\mu}\mY^{\hat\alpha}$ is not degenerate, $\mN^{\hat\alpha}$ is decided as following
\begin{eqnarray}
\mN_{\hat\alpha}=\frac{1}{d!}\mh^{-\frac12}\varepsilon^{\mu_1\cdots\mu_d} \epsilon_{\hat\alpha\hat\alpha_1\cdots\hat\alpha_{d+1}}\mD_{\mu_1} \mY^{\hat\alpha_1}\cdots \mD_{\mu_d} \mY^{\hat\alpha_d} \mY^{\hat\alpha_{d+1}}\,,
\end{eqnarray}
where
\begin{eqnarray}
\mh=-\det(\mh_{\mu\nu})\,,~~~~~~ \varepsilon^{01\cdots {d-1}}=1\,.
\end{eqnarray}
Besides the metric $\mh_{\mu\nu}$, another important $SO(2,d)$ invariant quantity on $\Sigma$ is
\begin{eqnarray}
\mK_{\mu\nu}=\mD_{\nu}\mY^{\hat\alpha}\mD_{\mu}\mN_{\hat\alpha}\,.
\end{eqnarray}
Obviously, it is related to the extrinsic curvature of $\Sigma$ in the usual geometric language. Analogy with the bulk results (\ref{bulkdev}), the derivatives on the basis $\{\mY,\mD_{\mu}\mY,\mN\}$ naturally implies the connection with torsion on $\Sigma$,
\begin{eqnarray}
\mD_{\mu}\mN^{\hat\alpha}\!\!\!\!&=&\!\!\!\!\mK_{\mu}{}^{\nu}\mD_{\nu} \mY^{\hat\alpha}\,,
\cr \mD_{\mu}\mD_{\nu}\mY^{\hat\alpha}\!\!\!\!&=&\!\!\!\! \ell^{-2}\mh_{\mu\nu}\mY^{\hat\alpha} +\mGamma^{\rho}{}_{\nu\mu}\mD_{\rho} \mY^{\hat\alpha}-\ell^{-2}\mK_{\mu\nu}\mN^{\hat\alpha}\,,
\end{eqnarray}
where
\begin{eqnarray}
\mGamma^{\mu}{}_{\nu\rho}\!\!\!\!&=&\!\!\!\!\hat\mGamma^{\mu}{}_{\nu\rho}+{\mt}^{\mu}{}_{\nu\rho}\,,
\cr
\hat\mGamma^{\mu}{}_{\mu_1\mu_2}
\!\!\!\!&=&\!\!\!\!\tfrac12\mH^{\mu\nu}\left(\md_{\mu_2}\mh_{\nu \mu_1}+\md_{\mu_1}\mh_{\nu\mu_2}
-\md_{\nu}\mh_{\mu_1\mu_2}\right)\,,
\cr
{\mt}^{\mu}{}_{\mu_1\mu_2}
\!\!\!\!&=&\!\!\!\!\frac{1}{2}\mH^{\mu\nu}\left[(\mF_{\nu\mu_1})^{\hat\beta_1\hat\beta_2}\mD_{\mu_2}\mY_{\hat\beta_1}
+(\mF_{\nu\mu_2})^{\hat\beta_1\hat\beta_2}\mD_{\mu_1}\mY_{\hat\beta_1} -(\mF_{\mu_1\mu_2})^{\hat\beta_1\hat\beta_2}\mD_{\nu}\mY_{\hat\beta_1}
\right]\mY_{\hat\beta_2}\,,~~~~~~~~
\end{eqnarray}
and the spacetime indices on $\Sigma$ are lowering and rising by $\mh_{\mu\nu}$ and its inverse $\mH^{\mu\nu}$.
The flux is decomposed as
\begin{eqnarray}
(\mF_{\mu_1\mu_2})^{\halpha\hbeta}
\!\!\!\!&=&\!\!\!\!\left[\mR^{\nu_1\nu_2}{}_{\mu_1\mu_2}
+2\ell^{-2}\left(\delta_{[\mu_1}^{\nu_1}\delta_{\mu_2]}^{\nu_2}-\mK_{[\mu_1}{}^{\nu_1}\mK_{\mu_2]}{}^{\nu_2}\right)\right]\mD_{\nu_1}\mY^{[\hat\alpha}\mD_{\nu_2}\mY^{\hat\beta]}
\cr\!\!\!\!&&\!\!\!\!-4\ell^{-4}\mK_{[\mu_1\mu_2]}\mY^{[\hat\alpha}\mN^{\hat\beta]} +4\ell^{-2}\mt^{\nu}{}_{[\mu_2\mu_1]}\mY^{[\hat\alpha}\mD_{\nu}\mY^{\hat\beta]}
\cr\!\!\!\!&&\!\!\!\!-4\ell^{-2}\left(\bcD_{[\mu_1}\mK_{\mu_2]}{}^{\nu}+\mt^{\nu_1}{}_{[\mu_2\mu_1]}\mK_{\nu_1}{}^{\nu}\right)\mN^{[\hat\alpha}\mD_{\nu}\mY^{\hat\beta]}
~~~~~~~~
\end{eqnarray}
where $\bcD_{\mu}$ and $\mR^{\nu_1\nu_2}{}_{\mu_1\mu_2}$ are respectively the covariant derivative and curvature tensor for the connection $\mGamma^{\nu}{}_{\rho\mu}$. It is also straightforward to check that
the Bianchi identity $\mD\mF=0$ is equivalent to the two Bianchi identities for Riemannian geometry.
\subsubsection{The canonical structure}
In terms of the above hypersurface notation, the bulk action becomes
\begin{eqnarray}
\!\!\!\!&&\!\!\!\!S[\mA,\mPhi, \mY]
\cr\!\!\!\!&=&\!\!\!\!
\frac1{g_0}\int \rd z\rd^d x\,\epsilon_{\hat\alpha_1\cdots\hat\alpha_{d+2}}\varepsilon^{\mu_1\cdots\mu_d} \Big\{\dot\mA_{\mu_1}^{\hat\alpha_{d+1}\hat\alpha_1}\mD_{\mu_2} \mY^{\hat\alpha_2}\cdots\mD_{\mu_d} \mY^{\hat\alpha_d} \mY^{\hat\alpha_{d+2}}
\cr&&~~~~~~~-\left[\frac{d-1}2(\mF_{\mu_1\mu_2})^{\hat\alpha_1\hat\alpha_2} -2\ell^{-2}\mD_{\mu_1} \mY^{\hat\alpha_1}\mD_{\mu_2} \mY^{\hat\alpha_2}\right] \mD_{\mu_3} \mY^{\hat\alpha_3}\cdots\mD_{\mu_d}  \mY^{\hat\alpha_d} \mY^{\hat\alpha_{d+1}}\dot \mY^{\hat\alpha_{d+2}}
\cr&&~~~~~~~
-\frac{d-1}4\ell^2 (\mF_{\mu_1\mu_2})^{\hat\alpha_1\hat\alpha_2}\mD_{\mu_3} \mY^{\hat\alpha_3}\cdots\mD_{\mu_d} \mY^{\hat\alpha_d}\mPhi^{\hat\alpha_{d+1}\hat\alpha_{d+2}}\Big\}
\end{eqnarray}
where the coupling constant is $g_0=2(-1)^d\,(d-1)!\kappa^2\ell$. For simplicity, we will set $g_0=1$ in the subsequent part of this paper. We notice that $\mPhi=A_z$ is a free Lagrangian multiplier which gives rise to exactly the constraint (\ref{mProjEin}).

The canonical momentums for the dynamical field $\mA$ and $\mY$ are given by\footnote{More rigorously, we should incorporate the constraint $\mY_{\alpha}\mY^{\alpha}=-\ell^2$ manifestly in the canonical procedure. It will lead to the secondary constraint  $\mY^{\alpha}\Pi_{\alpha}=0$ which will help us deciding $\Pi_{\alpha}$ unambiguously. }
\begin{eqnarray}\label{Pimu}
(\Pi^{\mu})_{\hat\alpha\hat\beta}\!\!\!\!&=&\!\!\!\!\frac{\partial{\mathcal L}}{\partial (\dot\mA_{\mu})^{\hat\alpha\hat\beta}}
=-\epsilon_{\hat\alpha\hat\beta\hat\alpha_1\cdots\hat\alpha_{d}} \varepsilon^{\mu\mu_2\cdots\mu_d} \mY^{\hat\alpha_1}\mD_{\mu_2} \mY^{\hat\alpha_2}\cdots\mD_{\mu_d} \mY^{\hat\alpha_d}\,,
\\\label{Pi}
\Pi_{\hat\alpha}\!\!\!\!&=&\!\!\!\!\frac{\partial{\mathcal L}}{\partial \dot \mY^{\hat\alpha}}
\cr\!\!\!\!&=&\!\!\!\!(-1)^d\epsilon_{\hat\alpha\hat\alpha_1\cdots\hat\alpha_{d+1}} \varepsilon^{\mu_1\cdots\mu_d}\left[\frac{d-1}2(\mF_{\mu_1\mu_2})^{\hat\alpha_1\hat\alpha_2} -\frac{2}{\ell^2}\mD_{\mu_1} \mY^{\hat\alpha_1}\mD_{\mu_2} \mY^{\hat\alpha_2}\right] \mD_{\mu_3} \mY^{\hat\alpha_3}\cdots\mD_{\mu_d}  \mY^{\hat\alpha_d} \mY^{\hat\alpha_{d+1}}
\,.~~~~~~~~~~
\end{eqnarray}
The corresponding Hamiltonian is simply the constraint for $\mPhi$
\begin{eqnarray}
H_0\!\!\!\!&=&\!\!\!\!\int_{\Sigma}\rd^d x\,[(\Pi^{\mu})_{\hat\alpha\hat\beta}(\dot\mA_{\mu})^{\hat\alpha\hat\beta}+\Pi_{\hat\alpha}\dot \mY^{\hat\alpha}-\mathcal L]
\cr\!\!\!\!&=&\!\!\!\! \frac{d-1}4\ell^2 \int_{\Sigma}\rd^d x\,\epsilon_{\hat\alpha_1\cdots\hat\alpha_{d+2}}\varepsilon^{\mu_1\cdots\mu_d} (\mF_{\mu_1\mu_2})^{\hat\alpha_1\hat\alpha_2}\mD_{\mu_3} \mY^{\hat\alpha_3}\cdots\mD_{\mu_d} \mY^{\hat\alpha_d}\mPhi^{\hat\alpha_{d+1}\hat\alpha_{d+2}}
\,.
\end{eqnarray}
Since that only first order $z$-derivative of $\mA$ and $\mY$ appeared in $\mathcal L$, the canonical momentums are decided by the canonical coordinates themselves instead of their radial derivatives. 
For such kind of constrained phase space, it is impossible to reproduce the expressions (\ref{Pimu}) and (\ref{Pi}) of the canonical momentums just from the Hamiltonian $H_0$ itself. Thus one can not to come back from $H_0$ to the original Lagrangian by performing the Legendre transformation backwardly. In fact, the Poisson bracket is also ill defined due to the mixing of the degree of freedom of the canonical pairs.
To solve these problems, it is better to incorporate the expressions (\ref{Pimu}) and (\ref{Pi}) manifestly as additional phase space constraints in the following Hamiltonian action\cite{Banados:2016zim}
\begin{eqnarray}
\!\!\!\!&&\!\!\!\!S[\mA_{\mu},\Pi^{\mu}, \mY,\Pi,\mPhi,\lambda_{\mu},\lambda]
\cr\!\!\!\!&=&\!\!\!\!\int\rd z\rd^d x\,\Big\{(\Pi^{\mu})_{\hat\alpha\hat\beta}(\dot\mA_{\mu})^{\hat\alpha\hat\beta}+\Pi_{\hat\alpha}\dot \mY^{\hat\alpha} -\frac{d-1}4\ell^2 \epsilon_{\hat\alpha\hat\beta\hat\alpha_1\cdots\hat\alpha_{d}}\varepsilon^{\mu_1\cdots\mu_d} (\mF_{\mu_1\mu_2})^{\hat\alpha_1\hat\alpha_2}\mD_{\mu_3} \mY^{\hat\alpha_3}\cdots\mD_{\mu_d} \mY^{\hat\alpha_d}\mPhi^{\hat\alpha\hat\beta}
\cr\!\!\!\!&&\!\!\!\!-(\lambda_{\mu})^{\hat\alpha\hat\beta}[(\Pi^{\mu})_{\hat\alpha\hat\beta}+\epsilon_{\hat\alpha\hat\beta\hat\alpha_1\cdots\hat\alpha_{d}} \varepsilon^{\mu\mu_2\cdots\mu_d} \mY^{\hat\alpha_1}\mD_{\mu_2} \mY^{\hat\alpha_2}\cdots\mD_{\mu_d} \mY^{\hat\alpha_d}]
\cr\!\!\!\!&&\!\!\!\!-\lambda^{\hat\alpha}\Big[\Pi_{\hat\alpha}-(-1)^d\epsilon_{\hat\alpha\hat\alpha_1\cdots\hat\alpha_{d+1}} \varepsilon^{\mu_1\cdots\mu_d}\left((d-1)(\mF_{\mu_1\mu_2})^{\hat\alpha_1\hat\alpha_2} -2\ell^{-2}\mD_{\mu_1} \mY^{\hat\alpha_1}\mD_{\mu_2} \mY^{\hat\alpha_2}\right) \mD_{\mu_3} \mY^{\hat\alpha_3}\cdots\mD_{\mu_d}  \mY^{\hat\alpha_d} \mY^{\hat\alpha_{d+1}}\Big] \Big\}
\,.~~~~~~~
\end{eqnarray}
Now the Poisson bracket can be well defined as usual
\begin{eqnarray}
\{P,Q\}=\int_{\Sigma}\rd^d x \left[\frac{\delta P}{\delta(\mA_{\mu})^{\hat\alpha\hat\beta}}
\frac{\delta Q}{\delta(\Pi^{\mu})_{\hat\alpha\hat\beta}}-\frac{\delta P}{\delta(\Pi^{\mu})_{\hat\alpha\hat\beta}}
\frac{\delta Q}{\delta(\mA_{\mu})^{\hat\alpha\hat\beta}}
+\frac{\delta P}{\delta \mY^{\hat\alpha}}
\frac{\delta Q}{\delta\Pi_{\hat\alpha}}-\frac{\delta P}{\delta\Pi_{\hat\alpha}}
\frac{\delta Q}{\delta \mY^{\hat\alpha}}\right]\,.
\end{eqnarray}

What is the physical meaning of the constraint (\ref{mProjEin})? By using (\ref{Pimu}) and (\ref{Pi}), we find that
\begin{eqnarray}
\!\!\!\!&&\!\!\!\!\bu^{\hat\alpha\hat\beta}[\mD_{\mu}(\Pi^{\mu})_{\hat\alpha\hat\beta}+\Pi_{[\hat\alpha} \mY_{\hat\beta]}]
\cr\!\!\!\!&=&\!\!\!\! \bu^{\hat\alpha\hat\beta}\epsilon_{\hat\alpha\hat\beta\hat\alpha_1\cdots\hat\alpha_{d}} \varepsilon^{\mu_1\mu_2\cdots\mu_d}\left[-\mD_{\mu_1} \mY^{\hat\alpha_1}\mD_{\mu_2} \mY^{\hat\alpha_2}\cdots\mD_{\mu_d} \mY^{\hat\alpha_d}+\frac{d-1}2(\mF_{\mu_1\mu_2})^{\hat\alpha_1\hat\gamma} \mY_{\hat\gamma} \mY^{\hat\alpha_2}\mD_{\mu_3} \mY^{\hat\alpha_3}\cdots\mD_{\mu_d} \mY^{\hat\alpha_d}\right]
\cr\!\!\!\!&&\!\!\!\! -\frac{(d-1)}{2}\epsilon_{\hat\alpha\hat\alpha_1\cdots\hat\alpha_{d+1}} \varepsilon^{\mu_1\cdots\mu_d}\bu^{\hat\alpha\hat\beta}(\mF_{\mu_1\mu_2})^{\hat\alpha_1\hat\gamma} \mY_{\hat\gamma} \mY^{\hat\alpha_{2}} \mD_{\mu_3} \mY^{\hat\alpha_3}\cdots\mD_{\mu_d}  \mY^{\hat\alpha_d}
\cr\!\!\!\!&&\!\!\!\!-\frac{(d-1)\ell^2}4\epsilon_{\hat\alpha\hat\beta\hat\alpha_1\cdots\hat\alpha_{d}} \varepsilon^{\mu_1\cdots\mu_d}\bu^{\hat\alpha\hat\beta}(\mF_{\mu_1\mu_2})^{\hat\alpha_1\hat\alpha_2} \mD_{\mu_3} \mY^{\hat\alpha_3}\cdots\mD_{\mu_d}  \mY^{\hat\alpha_d}
\cr\!\!\!\!&&\!\!\!\!+\epsilon_{\hat\alpha\hat\beta\hat\alpha_1\cdots\hat\alpha_{d}} \varepsilon^{\mu_1\cdots\mu_d}\bu^{\hat\alpha\hat\beta} \mD_{\mu_1} \mY^{\hat\alpha_1}\cdots\mD_{\mu_d} \mY^{\hat\alpha_d}
\cr\!\!\!\!&=&\!\!\!\! -\frac{(d-1)\ell^2}4\epsilon_{\hat\alpha\hat\beta\hat\alpha_1\cdots\hat\alpha_{d}} \varepsilon^{\mu_1\cdots\mu_d}\bu^{\hat\alpha\hat\beta}(\mF_{\mu_1\mu_2})^{\hat\alpha_1\hat\alpha_2} \mD_{\mu_3} \mY^{\hat\alpha_3}\cdots\mD_{\mu_d}  \mY^{\hat\alpha_d}
\,.
\end{eqnarray}
Therefore, once the expressions (\ref{Pimu}) and (\ref{Pi}) of canonical momentums are given, the equation  (\ref{mProjEin}) is indeed  equivalent to the constraint for the $SO(2,d)$ gauge invariance
\begin{eqnarray}
G_{0}[\bu]=\int_{\Sigma}\rd^d x\,\bu^{\hat\alpha\hat\beta}  [\mD_{\mu}(\Pi^{\mu})_{\hat\alpha\hat\beta}+\Pi_{[\hat\alpha} \mY_{\hat\beta]}]
\end{eqnarray}
which generates the $SO(2,d)$ transformation via the Poisson brackets
\begin{eqnarray}
\{(\mA_{\mu})^{\hat\alpha\hat\beta},G_{0}[\bu]\}\!\!\!\!&=&\!\!\!\!
-\mD_{\mu} \bu^{\hat\alpha\hat\beta}\,,~~~~~~
\{ \mY^{\hat\alpha},G_{0}[\bu]\}=
\bu^{\hat\alpha}{}_{\hat\beta} \mY^{\hat\beta}\,,
\cr  \{(\Pi^{\mu})_{\hat\alpha\hat\beta},G_{0}[\bu]\}\!\!\!\!&=&\!\!\!\!
2\bu_{[\hat\beta}{}^{\hat\gamma}(\Pi^{\mu})_{\hat\alpha]\hat\gamma}\,,
~~~~~ \{\Pi_{\hat\alpha},G_{0}[\bu]\}
=\bu_{\hat\alpha}{}^{\hat\gamma}\Pi_{\hat\gamma}\,.
\end{eqnarray}
Now, by linear redefinition of the free lagrangian multipliers, the Hamiltonian action
becomes
\begin{eqnarray}
\!\!\!\!&&\!\!\!\!S[\mA_{\mu},\Pi^{\mu}, \mY,\Pi,\mPhi,\lambda_{\mu},\lambda]
\cr\!\!\!\!&=&\!\!\!\!\int\rd z\rd^d x\,\Big\{(\Pi^{\mu})_{\hat\alpha\hat\beta}(\dot\mA_{\mu})^{\hat\alpha\hat\beta}+\Pi_{\hat\alpha}\dot \mY^{\hat\alpha} +\mPhi^{\hat\alpha\hat\beta}[\mD_{\mu}(\Pi^{\mu})_{\hat\alpha\hat\beta}+\Pi_{[\hat\alpha} \mY_{\hat\beta]}]
\cr\!\!\!\!&&\!\!\!\!-\lambda^{\hat\alpha}\Big[\Pi_{\hat\alpha}-(-1)^d\epsilon_{\hat\alpha\hat\alpha_1\cdots\hat\alpha_{d+1}} \varepsilon^{\mu_1\cdots\mu_d}\left(\frac{d-1}2(\mF_{\mu_1\mu_2})^{\hat\alpha_1\hat\alpha_2} -2\ell^{-2}\mD_{\mu_1} \mY^{\hat\alpha_1}\mD_{\mu_2} \mY^{\hat\alpha_2}\right) \mD_{\mu_3} \mY^{\hat\alpha_3}\cdots\mD_{\mu_d}  \mY^{\hat\alpha_d} \mY^{\hat\alpha_{d+1}}\Big]
\cr\!\!\!\!&&\!\!\!\!-(\lambda_{\mu})^{\hat\alpha\hat\beta}[(\Pi^{\mu})_{\hat\alpha\hat\beta}+\epsilon_{\hat\alpha\hat\beta\hat\alpha_1\cdots\hat\alpha_{d}} \varepsilon^{\mu\mu_2\cdots\mu_d} \mY^{\hat\alpha_1}\mD_{\mu_2} \mY^{\hat\alpha_2}\cdots\mD_{\mu_d} \mY^{\hat\alpha_d}]\Big\}
\cr\!\!\!\!&=&\!\!\!\!\int\rd z\rd^d x\,\Big\{(\Pi^{\mu})_{\hat\alpha\hat\beta}(\dot\mA_{\mu})^{\hat\alpha\hat\beta}+\Pi_{\hat\alpha}\dot \mY^{\hat\alpha} +\mPhi^{\hat\alpha\hat\beta}[\mD_{\mu}(\Pi^{\mu})_{\hat\alpha\hat\beta}+\Pi_{\hat\alpha} \mY_{\hat\beta}]
\cr&&~~~~~~~~~~~~-\lambda^{\hat\alpha}\Big[\Pi_{\hat\alpha}-3(-1)^d(d-1)!\mh^{\frac12}\, \mN_{[\hat\alpha}\mD^{\mu_1} \mY_{\hat\alpha_1}\mD^{\mu_2} \mY_{\hat\alpha_2]} (\mF_{\mu_1\mu_2})^{\hat\alpha_1\hat\alpha_2} +2(-1)^dd!\,\ell^{-2}\mh^{\frac12}\,\mN_{\hat\alpha}\Big]
\cr&&~~~~~~~~~~~~-(\lambda_{\mu})^{\hat\alpha\hat\beta}[(\Pi^{\mu})_{\hat\alpha\hat\beta} -(-1)^d2(d-1)!\,\mh^{\frac12}\,\mN_{[\hat\alpha}\mD^{\mu} \mY_{\hat\beta]}]\Big\}
~~~~~~~~~~~~~~~~~~~~~~~~
\end{eqnarray}
where we have simplified the expression by using the $\mN^{\hat\alpha}$ and $\mh$ defined previously.

Furthermore, in order to understand the physical meaning of other constraints, it is better to decompose $\lambda^{\hat\alpha}$
by the intrinsic $SO(2,d)$ basis{\footnote{Here we have used the condition $\lambda^{\hat\alpha}\mY_{\hat\alpha}=0$ which is inherited from $\Pi^{\hat\alpha}\mY_{\hat\alpha}=0$.}}
\begin{eqnarray}
\lambda^{\hat\alpha}\!\!\!\!&=&\!\!\!\!\ell^{-1}\fn\mN^{\hat\alpha}+\fn^{\mu}\mD_{\mu} \mY^{\hat\alpha}
\,.~~~~~
\end{eqnarray}
The corresponding constraint decouples as following
\begin{eqnarray}
G_{1}\!\!\!\!&=&\!\!\!\!\int_{\Sigma}\rd^d x \lambda^{\hat\alpha}\Big[\Pi_{\hat\alpha}-3(-1)^d(d-1)!\mh^{\frac12}\, \mN_{[\hat\alpha}\mD^{\mu_1} \mY_{\hat\alpha_1}\mD^{\mu_2} \mY_{\hat\alpha_2]} (\mF_{\mu_1\mu_2})^{\hat\alpha_1\hat\alpha_2} +2(-1)^dd!\,\ell^{-2}\mh^{\frac12}\,\mN_{\hat\alpha}\Big]
\cr\!\!\!\!&=&\!\!\!\!\int_{\Sigma}\rd^d x\,\Big\{\ell^{-1}\fn\Big[\Pi_{\hat\alpha}\mN^{\hat\alpha}
-(-1)^d(d-1)!\ell^2\mh^{\frac12}\, \mD^{\mu_1}\mY_{[\hat\alpha_1}\mD^{\mu_2}\mY_{\hat\alpha_2]} (\mF_{\mu_1\mu_2})^{\hat\alpha_1\hat\alpha_2}
+2(-1)^dd!\,\mh^{\frac12}\Big]
\cr&&~~~~~~~~~+\fn^{\mu}\Big[\Pi_{\hat\alpha}\mD_{\mu}\mY^{\hat\alpha}
-2(-1)^d(d-1)!\mh^{\frac12}\, \mN_{[\hat\alpha_1}\mD^{\mu_1}\mY_{\hat\alpha_2]} (\mF_{\mu_1\mu})^{\hat\alpha_1\hat\alpha_2} \Big]
\Big\}\,.~~~~~~
\end{eqnarray}
By using (\ref{Pimu}), the $\fn^{\mu}$ components can be further re-expressed as homogenous functions of the canonical momentums
\begin{eqnarray}
\!\!\!\!&&\!\!\!\!\Pi_{\hat\alpha}\mD_{\mu}\mY^{\hat\alpha}
-2(-1)^d(d-1)!\mh^{\frac12}\, \mN_{[\hat\alpha_1}\mD^{\mu_1}\mY_{\hat\alpha_2]} (\mF_{\mu_1\mu})^{\hat\alpha_1\hat\alpha_2}
=\Pi_{\hat\alpha}\mD_{\mu}\mY^{\hat\alpha}
+(\Pi^{\nu})_{\hat\alpha_1\hat\alpha_2}(\mF_{\mu\nu})^{\hat\alpha_1\hat\alpha_2}
\,.~~~~~~~~
\end{eqnarray}
Similarly, we can rewrite the $\fn$ component as
\begin{eqnarray}
\!\!\!\!&&\!\!\!\!\Pi_{\hat\alpha}\mN^{\hat\alpha}
-(-1)^d(d-1)!\ell^2\mh^{\frac12}\, \mD^{\mu_1}\mY_{[\hat\alpha_1}\mD^{\mu_2}\mY_{\hat\alpha_2]} (\mF_{\mu_1\mu_2})^{\hat\alpha_1\hat\alpha_2}
+2(-1)^dd!\,\mh^{\frac12}
\cr\!\!\!\!&=&\!\!\!\!
\Pi_{\hat\alpha}\mN^{\hat\alpha}+\ell^{-2}\partial_{\mu}\left[(\Pi^{\mu})_{\halpha\hbeta}\mY^{[\halpha}\mN^{\hbeta]}\right]
-(-1)^d(d-1)!\mh^{\frac12}\Big[\ell^2\mD^{\mu_1}\mY_{[\hat\alpha_1}\mD^{\mu_2}\mY_{\hat\alpha_2]} (\mF_{\mu_1\mu_2})^{\hat\alpha_1\hat\alpha_2}-2d\Big]\,.~~~~~~~~
\end{eqnarray}
Due to (\ref{Pimu}), the additional $\ell^{-2}\partial_{\mu}[(\Pi^{\mu})_{\halpha\hbeta}\mY^{[\halpha}\mN^{\hbeta]}]$  term  simply vanishes. The reason for including such a term will be illustrated later in Section 3.2.3.

Finally, we can  reformulate the Hamiltonian action as
\begin{eqnarray}
\!\!\!\!&&\!\!\!\!S[\mA_{\mu},\Pi^{\mu}, \mY,\Pi,\mPhi,\lambda_{\mu},\lambda]
=\int\rd z\rd^d x\,\Big\{(\Pi^{\mu})_{\hat\alpha\hat\beta}(\dot\mA_{\mu})^{\hat\alpha\hat\beta}+\Pi_{\hat\alpha}\dot \mY^{\hat\alpha} -\mathcal H\Big\}\,.~~~~~~~~~
\end{eqnarray}
where the corresponding Hamiltonian
\begin{eqnarray}\label{BulkHamiltonian}
\!\!\!\!&&\!\!\!\!H=\int_{\Sigma}\rd^d x\,\mathcal H
=G_0[-\mPhi]+G^{(0)}_{1}[\fn]+G^{(1)}_{1}[\fn^{\mu}]+G_{2}[\lambda_{\mu}]
\,,~~~~~~~~~~
\end{eqnarray}
is the summation of the constraints
\begin{eqnarray}\label{G0}
G_0[\mPhi]\!\!\!\!&=&\!\!\!\!\int_{\Sigma}\rd^d x\,\mPhi^{\hat\alpha\hat\beta}\left[\mD_{\mu}(\Pi^{\mu})_{\hat\alpha\hat\beta} +\Pi_{[\hat\alpha}\mY_{\hat\beta]}\right] \,,\\
\label{G10}
G^{(0)}_{1}[\fn]\!\!\!\!&=&\!\!\!\!\int_{\Sigma}\rd^d x\, \fn\Big\{\Pi_{\hat\alpha}\mN^{\hat\alpha}+\ell^{-2}\partial_{\mu}\left[(\Pi^{\mu})_{\halpha\hbeta}\mY^{[\halpha}\mN^{\hbeta]}\right]
\cr&&~~~~~~~~~~~~-(-1)^d(d-1)!\mh^{\frac12}\Big[\ell^2\mD^{\mu_1}\mY_{[\hat\alpha_1}\mD^{\mu_2}\mY_{\hat\alpha_2]} (\mF_{\mu_1\mu_2})^{\hat\alpha_1\hat\alpha_2}-2d\Big]\Big\}
\,,\\\label{G11}
G^{(1)}_{1}[\fn^{\mu}]\!\!\!\!&=&\!\!\!\!\int_{\Sigma}\rd^d x\,\fn^{\mu}\Big[\Pi_{\hat\alpha}\mD_{\mu}\mY^{\hat\alpha}
+(\Pi^{\nu})_{\hat\alpha_1\hat\alpha_2}(\mF_{\mu\nu})^{\hat\alpha_1\hat\alpha_2}\Big]
\,,\\\label{G2}
G_{2}[\lambda_{\mu}]
\!\!\!\!&=&\!\!\!\!\int_{\Sigma}\rd^d x\,(\lambda_{\mu})^{\hat\alpha\hat\beta}\left[(\Pi^{\mu})_{\hat\alpha\hat\beta} -(-1)^d2(d-1)!\,\mh^{\frac12}\,\mN_{[\hat\alpha}\mD^{\mu}\mY_{\hat\beta]}\right]
\,.
\end{eqnarray}
By computing the Poisson brackets among these constraints, it is shown that $G_{0}$, $G^{(0)}_{1}$ and $G^{(1)}_{1}$ are 1st class constraints.  This is a natural result since these are in fact the generator of the gauge symmetries in the theory. Obviously, $G_{0}$ generates the $SO(2,d)$ gauge transformation, while $G^{(1)}_{1}$ generates the gauge covariant diffeomorphism transformation\cite{Banados:2016zim} on $\Sigma$.  The generator $G^{(0)}_{1}$ is related to the bulk diffeomorphism transformations along the transverse direction of $\Sigma$. Therefore, $G^{(0)}_{1}$ contains the $\Pi$-independent anomaly term which arises due to the change of the hypersurface $\Sigma$ itself. On the other hand, the constraint $G_{2}$ does not relate to any gauge symmetry of the theory, and one can check it is indeed a 2nd class constraint. There is a interesting physical meaning of (\ref{G2}). That is, the conjugation momentum of the gauge field $\mA$ is given by the $SO(2,d)$ covariant area operator{\footnote {Here, the area means the size of a co-dimension one submanifold on $\Sigma$.}}on $\Sigma$. We conjecture that this formula can be understood as a local statement of the holographicity of quantum gravity.

\subsection{CFT conservation laws and holography}
Previously, we have shown that the pull back of the bulk EOM on $\Sigma$ is related to the $SO(2,d)$ conservation law in the Hamiltonian formalism. In this part, we sketch how to built up the same structure by the dual CFT arguments.
\subsubsection{The CFT $SO(2,d)$ formalism}
In the usual description of the CFT, the theory $S[\phi;\bmg]$ is defined upon the background metric $\bmg_{\mu\nu}$. Including the transformation of background metric, the classically action is invariant
under the $d$-dimensional  diffeomorphism Diff$_d$ as well as the Weyl transformation
\begin{eqnarray}\label{WeylTrans}
\bmg_{\mu\nu}\to \tilde\bmg_{\mu\nu}=e^{2\Omega}\bmg_{\mu\nu}\,,~~~~~~~~~~~~\phi\to \tilde\phi=e^{-\Delta\Omega}\phi\,.
\end{eqnarray}
For the flat metric $\bmg_{\mu\nu}=\eta_{\mu\nu}$ with $d>2$, the subgroup of Diff$_d\times$Weyl which keeps $\eta_{\mu\nu}$ intact is the rigid conformal group $SO(2,d)$. For $d=2$, the classical conformal group will be enhanced to the group of holomorphic maps which contains the rigid $SO(2,d)$ as a subgroup. Inspired by the bulk results, we expect that this theory could be equivalently described by the $SO(2,d)$ background field.

Due to the rigid conformal invariance of the action $S[\phi;\eta]$, the stress tensor
\begin{eqnarray}
\bmT^{\mu\nu}(\phi,\bmg)=\frac{1}{\sqrt{\bmg}}\frac{\delta S}{\delta \bmg_{\mu\nu}}
\end{eqnarray}
satisfies that
\begin{eqnarray}
\partial_{\mu}\bmT^{\mu\nu}=0\,,~~~~\bmT^{[\mu\nu]}=0\,,~~~~\bmT^{\mu}{}_{\mu}=0\,.~~~~
\end{eqnarray}
Equivalently, we can reformulate them as the $SO(2,d)$ conservation law
\begin{eqnarray}\label{UVcons}
\partial_{\mu}({\bmJ}^{\mu})^{\hat \alpha\hat\beta}=0\,.
\end{eqnarray}
In the above, the $SO(2,d)$ current is decided by the EM tensor as
\begin{eqnarray}\label{UVcurrents}
({\bmJ}^{\mu})^{\hat \alpha\hat\beta}=2\bmX^{[\hat\alpha}\partial_{\nu}\bmX^{\hat\beta]}\bmT^{\mu\nu}\,,~~~
\end{eqnarray}
where $\bmX^{\hat\alpha}$ is the background $SO(2,d)$ ruler field with the configuration
\begin{eqnarray}\label{CFTEG}
\bmX^{\hat\mu} = x^\mu  \,, ~~~
\bmX^{\hat d} =   \frac{\ell^2-\eta_{\mu\nu}x^{\mu}x^{\nu}}{2\ell}\, ,
~~~\bmX^{\hat\bullet} =  \frac{\ell^2+\eta_{\mu\nu}x^{\mu}x^{\nu}}{2\ell}\,.~~~~
\end{eqnarray}
This background ruler field $\bmX^{\hat\alpha}$ satisfies the similar relations as in the bulk analysis
\begin{eqnarray}\label{bfY^2}
\bmX^{\hat\alpha}\bmX_{\hat\alpha} = 0  \,, ~~~~~~~
\partial_{\mu}\bmX^{\hat\alpha}\partial_{\nu}\bmX_{\hat\alpha}=\eta_{\mu\nu}\,.~~~~
\end{eqnarray}
At this level, $\ell$ is just a constant with mass dimension $-1$. Naively, we can regard it as an auxiliary  ``typical scale'' of CFT.  As a consequence of global scale invariance, the explicit value of $\ell$ does not affect the property (\ref{bfY^2}) as well as the conservation law (\ref{UVcons}). It only appears in (\ref{CFTEG}) as a formal level parameter for the rigid conformal currents.

The CFT $SO(2,d)$ ruler $\bmX^{\hat\alpha}$ can be viewed as the boundary dual of the bulk ruler field $Y^{\hat\alpha}$. In fact, regarding $\bmX^{\hat\alpha}$ as a background primary with conformal weight $\Delta=-1$, we can apply the bulk-boundary relation suggested in \cite{Wang:2015qfa}
\begin{eqnarray}
 Y^{\hat\alpha}(x,z)\!\!\!\!&=&\!\!\!\! z^{-1}{}_0F_1\!\left(;\Delta-\tfrac{d}2+1;-\tfrac{\ell^2 z^2}{4}\Box\right)\bmX^{\hat\alpha}(x)
\end{eqnarray}
to generate the corresponding bulk configuration for $Y^{\hat\alpha}$.
Especially, the configuration (\ref{CFTEG}) will gives rise to the embedding gauge configuration (\ref{bulkEmG}) of $Y^{\hat\alpha}$. By fixing the formula (\ref{CFTEG}), the diffeomorphism+Weyl transformations on $\bmX^{\hat\alpha}$ can be mapped to the local $SO(2,d)/ISO(1,d-1)$ transformations which are decided up to the $ISO(1,d-1)$ subgroup leaving $\bmX^{\hat\alpha}$ intact. Especially, the rigid $SO(2,d)$ transformations are mapped to the rigid conformal transformations.

Generically, after performing the local $SO(2,d)$ transformations on $\bmX^{\hat\alpha}$, it is unavoidable to incorporate the corresponding background $SO(2,d)$ gauge field $\bmA_{\mu}$. Then the metric is given by the $SO(2,d)$ invariant quadratic form
\begin{eqnarray}
\bmg_{\mu\nu}=\bfD_{\mu}\bmX^{\hat\alpha}\bfD_{\nu}\bmX_{\hat\alpha}\,.~~~~
\end{eqnarray}
Obviously, after turning on $\bmA$, it covers the generic curved background metric configurations. As in the previous section, one can also incorporate the torsion by considering the generic intrinsic expansion of $\bfD_{\mu}\bfD_{\nu}\bmX^{\hat\alpha}$ . A crucial point is that the physical degree of freedom for $\bmA_{\mu}$ can not be totaly fixed by the $d$-dimensional background geometric data $\bmg_{\mu\nu}$ and $\bmt^{\rho}{}_{\nu\mu}$. For example, the conditions $\bmg_{\mu\nu}=\eta_{\mu\nu}$ and $\bmt^{\rho}{}_{\nu\mu}=0$ are not sufficient to fixing $\bmA_{\mu}$ to be the pure gauge configuration. To analyze it in general,  we can introduce $\bmV^{\hat\alpha}$ which satisfies
\begin{eqnarray}\label{bmV}
\bmV_{\hat\alpha}\bfD_{\mu}\bmX^{\hat\alpha}=0\,,~~~~~~~\bmV_{\hat\alpha}\bmV^{\hat\alpha}=0\,,
~~~~~~~\bmV_{\hat\alpha}\bmX^{\hat\alpha}=1\,,~~~~
\end{eqnarray}
to establish the intrinsic $SO(2,d)$ basis $\{\bmX^{\hat\alpha}, \bfD_{\mu}\bmX^{\hat\alpha}, \bmV^{\hat\alpha}\}$ for CFT.
Then the $SO(2,d)$ flux is decomposed as
\begin{eqnarray}
(\bmF_{\mu_1\mu_2})^{\halpha\hbeta}
\!\!\!\!&=&\!\!\!\!\left(\bmR^{\nu_1\nu_2}{}_{\mu_1\mu_2}
-4\delta_{[\mu_1}^{[\nu_1}\hmf_{\mu_2]}{}^{\nu_2]}\right)\bfD_{\nu_1}\bmX^{[\hat\alpha}\bfD_{\nu_2}\bmX^{\hat\beta]}
-4\hmf_{[\mu_1\mu_2]}\bmX^{[\hat\alpha}\bmV^{\hat\beta]}
\cr\!\!\!\!&&\!\!\!\!-4\bmt^{\nu}{}_{[\mu_1\mu_2]}\bfD_{\nu}\bmX^{[\hat\alpha}\bmV^{\hat\beta]}
-4\left(\bmnabla_{[\mu_1}\hmf_{\mu_2]}{}^{\nu}-\bmt^{\nu_1}{}_{[\mu_1\mu_2]}\hmf_{\nu_1}{}^{\nu}\right)\bmX^{[\hat\alpha}\bfD_{\nu}\bmX^{\hat\beta]}
\,.~~~~~
\end{eqnarray}
Now it is clear that the missing part is the additional background data
\begin{eqnarray}
\hmf_{\mu\nu}=\bfD_{\nu}\bmX_{\hat\alpha}\bfD_{\mu}\bmV^{\hat\alpha}\,.
\end{eqnarray}
What is the meaning of $\hmf_{\mu\nu}$ in the traditional CFT language? 
In the subsequent discussions,  we shall illustrate this point by considering the Weyl invariance of the quantum theory.
\subsubsection{Weyl transformation and the quantum metric}
At the quantum level, besides the classical action, one also need to consider the symmetry transformation of the path integral measure. For example, one can define the path integral measure by using the eigen functions
\begin{eqnarray}\label{Eigeneq}
\Box_{\bmg}\phi_{k}(x)=\bmg^{\mu\nu}\nabla_{\mu}\nabla_{\nu}\phi_{k}(x)=\lambda_k\phi_k(x)
\end{eqnarray}
with the normalization
\begin{eqnarray}\label{EigeneqNorm}
\langle\phi_{k'},\phi_{k}\rangle_{\bmg}=\int\rd^d x\sqrt{\bmg(x)}\phi_{k'}^*(x)\phi_k(x)=\delta^{d}(k-k')\,.
\end{eqnarray}
Via the corresponding eigen expansion
\begin{eqnarray}
\phi(x)=\sum_{k}a_k[\bmg]\phi_k(x)\,,
\end{eqnarray}
the path integral measure 
is defined as
\begin{eqnarray}
[\cD\phi]_{\bmg}=\prod_{k}\rd a_k[\bmg]\,.
\end{eqnarray}
Since the eigen equation (\ref{Eigeneq}) itself is a scalar under diffeomorphism
and the normalization (\ref{EigeneqNorm}) is diffeomorphism invariant, we simple have
\begin{eqnarray}
\tilde\phi_{k}(\tilde x)=\phi_k(x)\,,~~~~~~~
\tilde\phi(\tilde x)=\phi(x)=\sum_{k}a_k\phi_k(x)=\sum_{k}a_k\tilde\phi_k(\tilde x)\,.
\end{eqnarray}
Then the path integral measure is diffeomorphism invariant in the sense
\begin{eqnarray}
[\cD\tilde\phi]_{\tilde\bmg}=\prod_{k}\rd a_k[\tilde\bmg]=\prod_{k}\rd a_k[\bmg]=[\cD\phi]_{\bmg}\,,
\end{eqnarray}
and the classical diffeomorphism invariance is naturally inherited by the quantum theory.

Under the Weyl transformation (\ref{WeylTrans}), the eigen equation (\ref{Eigeneq}) transforms as
\begin{eqnarray}
[\Box_{\tilde\bmg}-\lambda_k](\ep^{-\Delta\Omega}\phi_{k})= \ep^{-(\Delta+2)\Omega}[\Box_{\bmg}-\ep^{-2\Omega}\lambda_k]\phi_{k}+\cdots
\end{eqnarray}
where $\dots$ denotes the non-tensor terms which contain the derivatives of $\Omega$. Unlike the diffeomorphism transformation, the eigen function of $\Box_{\tilde\bmg}$ can not be simply constructed from $\phi_{k}(x)$ when $\Omega$ is not a constant. As a result, under the generic Weyl transformations, the eigen expansion shall change drastically and the path integral measure will be quite different
\begin{eqnarray}
[\cD\phi]_{\bmg}=\prod_{k}\rd a_k[\bmg]\neq \prod_{k}\rd a_k[\tilde\bmg]=[\cD\phi]_{\tilde\bmg}\,.
\end{eqnarray}
Then the Weyl symmetry is anomalous at the quantum level
\begin{eqnarray}
Z[e^{2\Omega}\bmg]\!\!\!\!&=&\!\!\!\!\ep^{\ii\cA_{\rm W}[e^{2\Omega}\bmg;\bmg]}Z[\bmg]
\end{eqnarray}
since the Weyl anomaly $\cA_{\rm W}[e^{2\Omega}\bmg;\bmg]$ is not zero for generic $\Omega(x)$. 

To address the quantum Weyl invariance properly, it is better to introduce a local scale factor $\zeta(x)$ for which the Weyl transformation is naturally assigned as
\begin{eqnarray}
\zeta\to\tilde\zeta=e^{\Omega}\zeta\,.
\end{eqnarray}
Now one can improve the eigen equation as well as the normalization condition to be scale dependent
\begin{eqnarray}\label{EigeneqS}
\Box_{\mg,\zeta}\phi_{k}\!\!\!\!&=&\!\!\!\! \zeta^{-\Delta}\Box_{\mg}\left(\zeta^{\Delta}\phi_{k}\right)=\lambda_k\phi_k\,,
\cr
\langle\phi_{k'},\phi_{k}\rangle_{\mg,\zeta}\!\!\!\!&=&\!\!\!\!\int\rd^d x\sqrt{\mg}\zeta^{2\Delta}\phi_{k'}^*\phi_k=\delta^{d}(k-k')\,,
\end{eqnarray}
where the scale dependent ``quantum metric''
\begin{eqnarray}
\mg_{\mu\nu}(x,\zeta)=\zeta^{-2}\bmg_{\mu\nu}(x)
\end{eqnarray}
is Weyl invariant by construction.
It induces the scale dependent eigen expansion
\begin{eqnarray}
\phi(x)\!\!\!\!&=&\!\!\!\!\sum_{k}a_k[\mg,\zeta]\phi_k(x)\,,~~~~~~~~~
a_k[\mg,\zeta]=\int\rd^d x\sqrt{\mg}\zeta^{2\Delta}\phi_{k}^*(x)\phi(x)\,.
\end{eqnarray}
%
Due to the additional transformation of $\zeta$, the eigen functions simply transforms as
\begin{eqnarray}
\tilde\phi_k(x) =e^{-\Delta\Omega(x)}\phi_k(x)\,.
\end{eqnarray}
Thus $a_k[\mg,\zeta]$ is invariant under the Weyl transformation
\begin{eqnarray}
\tilde\phi(x)\!\!\!\!&=&\!\!\!\!e^{-\Delta\Omega}\phi(x)=e^{-\Delta\Omega}\sum_{k}a_k[\mg,\zeta]\phi_k(x) =\sum_{k}a_k[\mg,\zeta]\tilde\phi_k(x)\,.
\end{eqnarray}
As a result, the corresponding path integral measure is Weyl invariant in the sense
\begin{eqnarray}
[\cD\tilde\phi]_{\tilde\mg,\tilde\zeta}=\prod_{k}\rd a_k[\tilde\mg,\tilde\zeta]=\prod_{k}\rd a_k[\mg,\zeta]=[\cD\phi]_{\mg,\zeta}
\end{eqnarray}
and the classical Weyl invariance will be naturally inherited by the scale dependent partition function
\begin{eqnarray}
Z[\tilde\bmg;\tilde\zeta]\!\!\!\!&=&\!\!\!\!\int [\cD\tilde\phi]_{\tilde\mg,\tilde\zeta}\ep^{\ii S[\tilde\phi,\tilde\bmg]}=\int [\cD\phi]_{\mg,\zeta}\ep^{\ii S[\tilde\phi(\phi),\tilde\bmg]}=\int [\cD\phi]_{\mg,\zeta}\ep^{\ii S[\phi,\bmg]}=Z[\bmg;\zeta]\,.
\end{eqnarray}
Now the original Weyl anomaly $\cA_{\rm W}[e^{2\Omega}\bmg;\bmg]$ is related to the ratio between the partition functions defined at different scales. In term of the language of quantum states, the construction of path integral measure defines the vacuum state $|\Omega(\bmg)\rangle$ for the theory $S[\phi,\bmg]$. Since the construction of the path integral measure $[\cD\phi]_{\mg,\zeta}$ is scale dependent, the vacuum state shall also be refined by the energy scale
\begin{eqnarray}
|\Omega(\bmg)\rangle&\to&|\Omega(\bmg;\zeta)\rangle\,.
\end{eqnarray}
The existence of Weyl anomaly means that the classical Weyl invariance is spontaneously broken by the refined structure of vacuum states.

%
In the above construction, the quantum metric $\mg_{\mu\nu}$ which is related to the classical metric $\bmg_{\mu\nu}$ by a local re-scaling.
More generically, one can consider the case where the path integral measure is defined by a quantum metric $\mh_{\mu\nu}(x, \zeta)$ which is totally different from the classical metric $\bmg_{\mu\nu}$. As the generalization of $\mg_{\mu\nu}$, we requires that $\mh_{\mu\nu}$ is invariant under the Weyl transformation. Then the corresponding path integral measure is Weyl invariant in the sense
\begin{eqnarray}
[\cD\tilde\phi]_{\tilde\mh,\tilde\zeta}\!\!\!\!&=&\!\!\!\!\prod_{k}\rd a_k[\tilde\mh,\tilde\zeta]=\prod_{k}\rd a_k[\mh,\zeta]=[\cD\phi]_{\mh,\zeta}\,.
\end{eqnarray}
The differences between the corresponding quantum metric and the classical metric 
will naturally give rise to the additional 2-tensor $\hmf_{\mu\nu}$
\begin{eqnarray}
\ell^{2}\hmf_{\mu\nu}\!\!\!\!&\sim&\!\!\!\!\zeta^{-2}\bmg_{\mu\nu}-\mh_{\mu\nu}+\cdots\,.
\end{eqnarray}
In terms of the language of quantum states, we have defined the bi-geometrical states $|\Omega(\bmg;\mh,\zeta)\rangle$. Especially, when the quantum metric is in the neighborhood of the classical one $\mh=\mg+O(\epsilon)$,
the states $|\Omega(\bmg;\mg+O(\epsilon),\zeta)\rangle$ can be viewed as the ``geometrical coherent states'' around the ``true vacuum'' $|\Omega(\bmg;\mg,\zeta)\rangle=|\Omega(\bmg;\zeta)\rangle$ at the scale $\zeta$. Measured by the regularized stress tensor operator $\bmT^{\mu\nu}_{\bmg}$ of the ``true vacuum'', these coherent states have non-zero expectation value $\langle\bmT^{\mu\nu}_{\bmg}\rangle\neq0$. 

Similarly, the torsion $\bmt^{\mu}{}_{\nu\rho}$ shall arise naturally by more careful treatment of the connection in the construction of path integral measure. The general partition function becomes
\begin{eqnarray}
Z[\bmg;\mh,\bmt,\zeta]\!\!\!\!&=&\!\!\!\!\int [\cD\phi]_{\mh,\bmt,\zeta}\ep^{\ii S[\phi,\bmg]}\,.
\end{eqnarray}
By reversing the procedure in Section 3.2.1, we can covert the quantum background geometry $\{\bmg,\hmf,\bmt\}$ to the background $SO(2,d)$ ruler and gauge fields $\{\bmX,\bmA\}$.
Acting on $\{\bmX,\bmA\}$, the Weyl transformation is given by
\begin{eqnarray}\label{UpWeyl}
\bmX^{\halpha}&\to& \tilde\bmX^{\halpha}=e^{\Omega}\bmX^{\halpha}\,,~~~~~~~~~~~~
\cr\bmA^{\halpha}{}_{\hbeta}&\to& (\tilde\bmA_{\mu})^{\halpha}{}_{\hbeta}=(\bmA_{\mu})^{\halpha}{}_{\hbeta} -\partial_{\mu}\Omega\left(\bmX^{\halpha}\bmV_{\hbeta}-\bmV^{\halpha}\bmX_{\hbeta}\right)\,.
\end{eqnarray}
Since the quantity $\bmV^{\halpha}$ is implicitly decided by $\{\bmX,\bmA\}$ in (\ref{bmV}), its Weyl transformation can be deduced from (\ref{UpWeyl}). The result is simply
\begin{eqnarray}
\bmV^{\halpha}~~\to~~\tilde\bmV^{\halpha}\!\!\!\!&=&\!\!\!\!e^{-\Omega}\bmV^{\halpha} \,.
\end{eqnarray}
\subsubsection{Bulk reconstruction}
Given the CFT ruler and gauge fields $\{\bmX,\bmA\}$, one can construct the bulk ruler and gauge fields  $\{\mY,\mA\}$ as
\begin{eqnarray}
\mY^{\hat\alpha}\!\!\!\!&=&\!\!\!\!\zeta^{-1}\bmX^{\hat\alpha}-\frac{\ell^2}2\zeta\bmV^{\hat\alpha}\,,
\cr (\mA_{\mu})^{\halpha}{}_{\hbeta}\!\!\!\!&=&\!\!\!\!(\bmA_{\mu})^{\halpha}{}_{\hbeta} +\zeta^{-1}\partial_{\mu}\zeta\left(\bmX^{\halpha}\bmV_{\hbeta}-\bmV^{\halpha}\bmX_{\hbeta}\right)\,.
\end{eqnarray}
According to (\ref{mN}), we further get
\begin{eqnarray}
\mN^{\hat\alpha}\!\!\!\!&=&\!\!\!\!\zeta^{-1}\bmX^{\hat\alpha}+\frac{\ell^2}2\zeta\bmV^{\hat\alpha}\,.
\end{eqnarray}
By introducing the scale $\zeta$ dependence, the above quantities are constructed to be Weyl invariant. Therefore, the infinitesimal Weyl transformation acting on the physical fields $\delta_{\rm W}\bmY,\delta_{\rm W}\bmA$ must compensate with the corresponding transformation $\delta\zeta=\omega\zeta$ acting on $\zeta$.
That is
\begin{eqnarray}\label{mWeyl}
\delta_{\rm W}\mY^{\halpha}\!\!\!\!&=&\!\!\!\!-\delta_{\zeta}\mY^{\halpha} =\omega(\zeta^{-1}\bmX^{\hat\alpha}+\frac{\ell^2}2\zeta\bmV^{\hat\alpha})=\omega\mN^{\halpha}\,,
\cr
\delta_{\rm W}(\mA_{\mu})^{\halpha\hbeta}\!\!\!\!&=&\!\!\!\!-\delta_{\zeta}(\mA_{\mu})^{\halpha\hbeta} =-2\partial_{\mu}\omega\bmX^{[\halpha}\bmV^{\hbeta]}
=-2\ell^{-2}\partial_{\mu}\omega\mY^{[\halpha}\mN^{\hbeta]}\,.
\end{eqnarray}

Now the scale dependent partition function becomes
\begin{eqnarray}
Z[\bmg;\mh,\bmt,\zeta]~~~\to~~~ Z[\bmX,\bmA,\zeta]~~~\to~~~Z[\mY,\mA,\zeta]= e^{\ii W[\mY,\mA,\zeta]}
\end{eqnarray}
where $W[\mA,\mY,\zeta]$ is the effective action at the local scale $\zeta$. From the CFT point of view, the relation between the bulk canonical pairs can be regarded as the source-response relation for the background source $\mY$ and $\mA_{\mu}$ on the state $|\Omega(\mY,\mA,\zeta)\rangle$. That is
\begin{eqnarray}\label{S-R}
\Pi\!\!\!\!&=&\!\!\!\!\mJ_{\rm cov}=\left\langle\Omega(\mY,\mA,\zeta)\left|\frac{\delta S}{\delta \mY}\right|\Omega(\mY,\mA,\zeta)\right\rangle\,,~~~~~~
\cr \Pi^{\mu}\!\!\!\!&=&\!\!\!\!\mJ^{\mu}_{\rm cov}=\left\langle\Omega(\mY,\mA,\zeta)\left|\frac{\delta S}{\delta \mA_{\mu}}\right|\Omega(\mY,\mA,\zeta)\right\rangle\,.
\end{eqnarray}
The subscript ``cov'' refers to the CFT covariant currents which are covariant under the $SO(2,d)$ gauge transformation. Obviously, the functional derivatives in defining the covariant currents only encounter the $\{\mY,\mA_{\mu}\}$ dependence in the classical action $S[\phi,\mY,\mA,\zeta]$. Due to the additional $\{\mY,\mA_{\mu}\}$ dependence in the path integral measure $[\cD\phi]_{\mY,\mA,\zeta}$, the covariant currents do not satisfy the Wess-Zumino consistent condition. In fact,  the Wess-Zumino consistent condition is satisfied by the consistent currents
\begin{eqnarray}\label{S-R}
\mJ_{\rm con}=\frac{\delta W[\mY,\mA,\zeta]}{\delta\mY}\,,~~~~~~~~\mJ^{\mu}_{\rm con}=\frac{\delta W[\mY,\mA,\zeta]}{\delta\mA_{\mu}}\,,
\end{eqnarray}
which are non-longer $SO(2,d)$ covariant in general.

Since the bulk and boundary theory are characterized by the same symmetries, we can naturally identify the CFT conservation laws with the 1st class constraints obtained in the bulk Hamiltonian analysis. The internal $SO(2,d)$ gauge transformation is
\begin{eqnarray}
\delta_{\rm int}(\mA_{\mu})^{\hat\alpha\hat\beta}=-\mD_{\mu}u^{\hat\alpha}{}_{\hat\beta}\,,~~~~~~~~
\delta_{\rm int}\mY^{\hat\alpha}=u^{\hat\alpha}{}_{\hat\beta}\mY^{\hat\beta} \,.
\end{eqnarray}
The corresponding conservation law
\begin{eqnarray}\label{mSO2d}
\mD_{\mu}(\mJ_{\rm cov}^{\mu})_{\hat\alpha\hat\beta}+(\mJ_{\rm cov})_{[\hat\alpha} \mY_{\hat\beta]}=0
\end{eqnarray}
matches with the $G_0$ constraint (\ref{G0}) explicitly. The $d$-dimensional gauge covariant diffeomorphism on $\Sigma$ is given by
\begin{eqnarray}
\delta_{\rm diff}(\mA_{\mu})^{\hat\alpha\hat\beta}=-\delta x^{\nu}(\mF_{\nu\mu})^{\hat\alpha\hat\beta}\,,~~~~~~~~
\delta_{\rm diff}\mY^{\hat\alpha}=-\delta x^{\mu}\mD_{\mu}\mY^{\hat\alpha} \,.
\end{eqnarray}
The corresponding conservation law
\begin{eqnarray}\label{mDiff}
(\mJ_{\rm cov})_{\hat\alpha}\mD_{\mu}\mY^{\hat\alpha}+(\mJ_{\rm cov}^{\nu})_{\hat\alpha\hat\beta}(\mF_{\mu\nu})^{\hat\alpha\hat\beta}=0 \,,
\end{eqnarray}
recovers the $G^{(1)}_{1}$ constraint (\ref{G11}). Given $\mJ_{\rm cov}^{\mu}$, (\ref{mDiff}) decides the $\mD_{\mu}\mY_{\hat\alpha}$ components of $(\mJ_{\rm cov})^{\hat\alpha}$. The $\mN^{\hat\alpha}$ component of $\mJ_{\rm cov}$ shall be fixed by the anomalous conservation law for Weyl transformation. From (\ref{mWeyl}), we get
\begin{eqnarray}\label{mWeyl}
(\mJ_{\rm cov})_{\hat\alpha}\mN^{\hat\alpha}+2\ell^{-2}\partial_{\mu}\left[(\mJ^{\mu}_{\rm cov})_{\halpha\hbeta}\mY^{[\halpha}\mN^{\hbeta]}\right]=\cA^{\rm cov}_{\rm W}\,,
\end{eqnarray}
where $\cA^{\rm cov}_{\rm W}$ is the covariant formula of the Weyl anomaly.
The corresponding bulk $G^{(0)}_{1}$ constraint (\ref{G10})  suggests that the covariant Weyl anomaly is
\begin{eqnarray}
\cA^{\rm cov}_{\rm W}
=(-1)^d(d-1)!\mh^{\frac12}\Big[\ell^2\mD^{\mu_1}\mY_{[\hat\alpha_1}\mD^{\mu_2}\mY_{\hat\alpha_2]} (\mF_{\mu_1\mu_2})^{\hat\alpha_1\hat\alpha_2}-2d\Big]\,,
\end{eqnarray}
for the dual CFT of Einstein gravity. On the other hand, given the explicit CFT, we can compute $\cA^{\rm cov}_{\rm W}$ directly by the DeWitt-Schwinger Method.

Finally, how to understand the bulk $G_2$ constraint (\ref{G2}) which fixes the expression for $\Pi_{\mu}$? Since $G_2$ is a 2nd class constraint, one can not derive it from the symmetry of the CFT. Luckily, we do have a chance to introduce it by hand at the CFT side.  During the standard QFT renormalization procedure, in order to decide the explicit form of the counter terms in the bare action, one should impose a renormalization condition for each background parameter. In this logic, a renormalization condition  should also be imposed for the CFT background metric. Equivalently, in the $SO(2,d)$ formalism, one need a renormalization condition for the background $SO(2,d)$ gauge field. For the CFT  which is dual to Einstein gravity, the renormalization condition to be imposed is just the $SO(2,d)$ covariant area law
\begin{eqnarray}\label{ReNC}
\mJ^{\mu}_{\rm cov}=\left\langle\Omega(\mY,\mA,\zeta)\left|\frac{\delta S}{\delta \mA_{\mu}}\right|\Omega(\mY,\mA,\zeta)\right\rangle=(-1)^d2(d-1)!\,\mh^{\frac12}\,\mN_{[\hat\alpha}\mD^{\mu}\mY_{\hat\beta]}\,.
\end{eqnarray}
In other words, at the renormalization scale $\zeta$, the renormalized physical area is defined by the expectation value of the $SO(2,d)$ current $\mJ^{\mu}=\frac{\delta S}{\delta \mA_{\mu}}$. As we mentioned earlier, this formula could be understood as a local statement of the holographicity of quantum gravity. It is very curious to clarify its relation with the other formulae of holographic area law, such as the Bekenstein-Hawking entropy, the Ryu-Takayanagi  formula\cite{Ryu:2006bv}, etc.

The renormalization condition (\ref{ReNC}) decides the explicit expression of the current $\mJ^{\mu}_{\rm cov}$. Then by applying the (anomalous-)conservation laws (\ref{mDiff}) and (\ref{mWeyl}), the explicit expression of the current $\mJ_{\rm cov}$ is also fixed. Substituting the expressions of $\mJ^{\mu}_{\rm cov}$ and $\mJ_{\rm cov}$ into (\ref{mSO2d}), we find the $SO(2,d)$ conservation law is anomalous
\begin{eqnarray}
\cA^{\rm cov}_{SO(2,d)}\!\!\!\!&=&\!\!\!\!\mD_{\mu}(\mJ_{\rm cov}^{\mu})_{\hat\alpha\hat\beta}+(\mJ_{\rm cov})_{[\hat\alpha} \mY_{\hat\beta]}
\cr\!\!\!\!&=&\!\!\!\!-\frac{(d-1)\ell^2}4\epsilon_{\hat\alpha\hat\beta\hat\alpha_1\cdots\hat\alpha_{d}} \varepsilon^{\mu_1\cdots\mu_d}(\mF_{\mu_1\mu_2})^{\hat\alpha_1\hat\alpha_2} \mD_{\mu_3} \mY^{\hat\alpha_3}\cdots\mD_{\mu_d}  \mY^{\hat\alpha_d}\,.
\end{eqnarray}
However, our construction of CFT itself is based on the exact background $SO(2,d)$ gauge invariance. Therefore, unlike the Weyl anomaly, the $SO(2,d)$ anomaly implies an inconsistency of the theory. At the quantum level, a consistent CFT can only be defined on the background where the covariant $SO(2,d)$ anomaly $\cA^{\rm cov}_{SO(2,d)}$ is vanishing at any given local scale $\zeta(x)$. Since $\cA^{\rm cov}_{SO(2,d)}$ equals to the pullback of bulk Einstein equation on $\Sigma=\{(x,z)|z=\zeta(x)\}$, it means that the RG flow of the consistent CFT background geometry should satisfies the full bulk Einstein equation. In this procedure, the bulk dynamics is emergent from the boundary $SO(2,d)$ non-anomalous condition. Alternatively, one can derive the radial components of Einstein equation directly as the Callan-Symanzik equation in our frame work.
\subsection{Generalizations}
\subsubsection{General gravity theories}
So far, we have shown that the bulk dynamics of pure Einstein gravity with negative cosmological constant is emergent from the vanishing of boundary $SO(2,d)$ covariant anomaly. On the other hand, there are various generalization of pure gravity theories which also contains AdS space as a vacuum solution. It is also possible to reformulate the 1st order formalism of these theories as $SO(2,d)$ gauge theories. The simplest example is the 2-dimensional Jackiw-Teitelboim dilaton gravity. In the $SO(2,1)$ gauge theory formalism, the action of JT gravity is
\begin{eqnarray}\label{JT}
\!\!\!\!&&\!\!\!\!S_{\rm JT}=\frac1{2\kappa^2}\int_{M} \epsilon_{\halpha_1\halpha_{2} \halpha_{3}} F^{\halpha_1\halpha_2}\Phi^{\halpha_3}\,,
\end{eqnarray}
where $F^{\halpha_1\halpha_2}$ is the field strength of the $SO(2,1)$ gauge field  and $\Phi^{\halpha}$ is the $SO(2,1)$ uplifting of the dilaton. As in the Chern-Simons action for the AdS$_3$ Einstein gravity, we need to implicitly impose the ruler field $Y^{\halpha}$ for constructing the spacetime metric
\begin{eqnarray}
g_{MN}\!\!\!\!&=&\!\!\!\!\rD_{M}Y^{\hat\alpha}\rD_{N}Y_{\hat\alpha}\,.
\end{eqnarray}
Expanding $\Phi^{\halpha}$ on the intrinsic basis, we have
\begin{eqnarray}
\Phi^{\hat\alpha}=\phi Y^{\hat\alpha}+\phi^{M}\rD_{M}Y^{\hat\alpha}
\end{eqnarray}
where $\phi$ is the usual dilaton field in the 2nd order formalism. In fact, an equivalent formula\cite{Isler:1989hq,Chamseddine:1989yz} of (\ref{JT}) has already been used in the exact quantization of JT gravity\cite{Iliesiu:2019xuh}.

More systematically, one can classify the $SO(2,d)$ gauge theories of gravity  by the characteristic class $I_{d+2}$ of the boundary anomaly \cite{Wang:2021ztz}. According to the descendant structure of anomalies,  the characteristic class  is given by the exterior derivative of the bulk action. For the JT gravity (\ref{JT}),  it is obviously
\begin{eqnarray}
\!\!\!\!&&\!\!\!\!I_{3}=\frac1{2\kappa^2}\epsilon_{\halpha_1\halpha_{2} \halpha_{3}} F^{\halpha_1\halpha_2}\wedge\rD\Phi^{\halpha_3}\,.
\end{eqnarray}
For the Einstein gravity (\ref{GRGT}), the computations in the Appendix imply that
\begin{eqnarray}\label{Chara}
I_{d+2}\!\!\!\!&=&\!\!\!\!\frac{(-1)^{d+1}\ell}{8\kappa^2(d-2)!} \epsilon_{\halpha_0\cdots\halpha_{d+1}}F^{\halpha_0\halpha_1}\wedge F^{\halpha_2\halpha_3}\wedge\rD Y^{\halpha_4}\wedge\cdots\wedge\rD Y^{\halpha_{d+1}}\,.
\end{eqnarray}
For $d=2$, it is just the Euler class for 4-dimensional manifolds. As the generalization of (\ref{Chara}), the characteristic classes for the Lovelock type of gravities is presented in \cite{Wang:2021ztz}
\begin{eqnarray}\label{CharaLove}
I_{d+2}\!\!\!\!&=&\!\!\!\!\sum_{i=1}^{\lfloor \frac{d}2\rfloor+1}a_i\epsilon_{\halpha_0\cdots\halpha_{d+1}}F^{\halpha_0\halpha_1}\wedge\cdots\wedge F^{\halpha_{2i-2}\halpha_{2i-1}}\wedge\rD Y^{\halpha_{2i}}\wedge\cdots\wedge\rD Y^{\halpha_{d+1}}
\,.
\end{eqnarray}
When $d$ is even, the last term with $i=\frac{d}2+1$ is again the Euler class for $(d+2)$-dimensional manifolds.
Therefore, (\ref{CharaLove}) shall be viewed as a generalization of the Euler class.

\subsubsection{Excited states and matter coupling}
In the previous discussions, we concentrated on the holographic emergence of the vacuum Einstein equation.  Under the general considerations, the bulk gravity also couples to various matter fields $\{\phi_i\}$, and the Einstein equation should include the matter stress tensor as sources. In the CFT point of view, the non-zero configurations of bulk matter field $\phi_i$ are related to the condensation of the dual CFT primary operator $\mathcal O_i$. In these cases, the dual CFT state $|\psi_i\rangle$ has non-zero expectation  $\langle\psi_i|\mathcal O_i|\psi_i\rangle$ for the operator $\mathcal O_i$. Therefore, $|\psi_i\rangle$ is no longer the CFT vacuum state but a generic excited state. Unlike the vacuum states $|\Omega(\bmg;\mh,\zeta)\rangle$, the $SO(2,d)$ transformations of excited states can not be absorbed by the transformations of $\mA$ and $\mY$. Instead, $|\psi_i\rangle$ transforms nontrivially as
\begin{eqnarray}
\delta|\psi_i\rangle=\epsilon \sum_{j}M_{ij}|\psi_j\rangle\,,
\end{eqnarray}
where the matrix $M_{ij}$ shall be decided by computing the corresponding OPE.
Then the  $SO(2,d)$ covariant conservation law should be modified as
\begin{eqnarray}
\left\langle\psi_i\left|\mD_{\mu}\frac{\delta S}{\delta (\mA_{\mu})^{\hat\alpha\hat\beta}}+\frac{\delta S}{\delta \mY^{[\hat\alpha}} \mY_{\hat\beta]}\right|\psi_i\right\rangle=J_{\psi_i}\,,
\end{eqnarray}
where the additional source
\begin{eqnarray}
J_{\psi_i}=\sum_{j}\Big(\langle\psi_i|M_{ij}|\psi_j\rangle+\langle\psi_j|(M^{\dagger})_{ji}|\psi_i\rangle\Big)
\end{eqnarray}
comes from the $SO(2,d)$ transformation of the excited state itself.
The quantum consistency of the $SO(2,d)$ background gauge field description further implies that
\begin{eqnarray}
\cA^{\rm cov}_{SO(2,d)}=J_{\psi_i}\,.
\end{eqnarray}
Translated to the bulk side, it means that the non-zero configuration of bulk matter fields will contribute as the source of the vacuum Einstein equation. This is the holographic origin of the matter fields back-reaction to the spacetime geometry. In principle, by computing $M_{ij}$ in CFT, one can decide how the matter fields couple to  gravity in the dual bulk theory.

Meanwhile, the bulk equation of motion for the matter field $\phi_i$ is constrained by the $SO(2,d)$ transformation of the dual CFT primary operators $\mathcal O_i$. By including the $SO(2,d)$ transformation of the energy scale, the finite scale primary condition will restrict the possible configurations of the bulk field. Especially, at the leading order, the finite scale primary condition requires that the bulk field  must satisfy the bulk free field equation \cite{Wang:2015qfa}. For example, the finite scale scalar primary in the pure AdS vacuum is given by
\begin{eqnarray}
\mathcal{O}(z,x)={z^{\Delta}}{}_0F_1\! \left(;\Delta-\tfrac{d}2+1;-\tfrac{z^2}{4}\Box\right)\mathcal{O}(x)\,.
\end{eqnarray}
This is exactly the normalizable solution of the bulk EOM for the dual scalar field $\phi$ in pure AdS background. For general backgrounds, we need to establish the most general finite scale primary condition which incorporates the contribution of $\hmf_{\mu\nu}$.

In supergravities and higher spin gauge theories, the bulk gravity field and matter fields are unified by the larger symmetry group $\mathcal{G}$ which contains $SO(2,d)$ as a subgroup. For these theories, one can still decide how the matter fields couple to the gravity in the ``perturbative'' way suggested above. In such a treatment, the larger symmetry structures shall arise order by order since our starting point is only the subgroup $SO(2,d)$.  Obviously, a better treatment is to start with the larger group $\mathcal{G}$ directly. Now the bulk theory must be a gauge theory with the gauge group $\mathcal{G}$. Of course, it contains the $SO(2,d)$ gauge theory of gravity as a sub-sector. On the other hand, the generic boundary dual theory should be defined based on the background gauge field $\mA$ with the same gauge group $\mathcal{G}$. Given a background configuration of $\mA$, the covariant conservation law for the $\mathcal{G}$ transformation could be anomalous
\begin{eqnarray}
\mD \mJ^{\rm cov}+\cdots=\cA^{\rm cov}_{\mathcal{G}}
\end{eqnarray}
where $\cdots$ denotes the contribution from the background ruler field. The quantum consistency of the background gauge field description requires that
\begin{eqnarray}
\cA^{\rm cov}_{\mathcal{G}}=0\,.
\end{eqnarray}
The bulk dynamics of the $\mathcal{G}$ gauge theory is emergent when the anomaly vanishing condition is imposed on any local energy scales. The explicit formula of the bulk $\mathcal{G}$ gauge theory shall be decided by the characteristic class of the boundary $\mathcal{G}$ anomaly.

\section{Summary}
In this paper, we apply a manifestly $SO(2,d)$ covariant formalism to explore the holographic emergency of the bulk dynamics. The bulk  gravity is reformulated as a $SO(2,d)$ gauge theory. The analysis in the Hamiltonian formalism shows that the bulk EOM is governed by the hypersurface $SO(2,d)$  conservation law. Providing the $SO(2,d)$ covariant conservation law is non-anomalous on arbitrary hypersurface $\Sigma$, the full bulk Einstein equation will be automatically implied.

For the dual CFT, a generic $SO(2,d)$ background fields description can be parallelly established by distinguishing the classical metric with the quantum metric which defines the path integral measure. Under this framework, the CFT covariant currents are naturally related to the bulk canonical momentums. By considering the local $SO(2,d)$, diffeomorphism and Weyl transformations in the dual CFT, the conservation laws in the bulk Hamiltonian formalism are easily recovered. Meanwhile, a $SO(2,d)$ covariant area law should be imposed as the renormalization condition of the background $SO(2,d)$ gauge field. Based on these constructions, 
the non-anomalous condition for the CFT $SO(2,d)$ gauge invariance gives rise to the dynamics of bulk geometry.
By considering the characteristics class of anomalies as well as the background gauge transformations of excited states, our construction can be generalized to a large class of bulk theories.

The $SO(2,d)$ gauge theory formulation of AdS gravity can be easily extended to the asymptotically dS or flat case. The only difference is  the gauge group becomes $SO(1,d+1)$ or $ISO(1,d)$. In principle, a similar holographic description of the corresponding gravity theory could be established by following the same procedures in this paper. As a subsequent task, it is curious enough to consider these generalizations in full details and comparing our procedure with the other proposals \cite{Strominger:2001pn,Bagchi:2014iea}. On the other hand, the Hamiltonian formalism obtained in this paper shows some analogy with the Hamiltonian based on the Ashtekar variables in loop quantum gravity\cite{Ashtekar:1986yd,Ashtekar:2004eh}. It is interesting to consider whether one can borrow the technics in LQG to perform the canonical quantization of this bulk $SO(2,d)$ gauge theory. 

\vskip 5mm

\vskip 1cm
\centerline{\bf\large Acknowledgments}
\vskip 5mm\noindent
The authors thank Chi-Ming Chang, Chen-Xu Han, Mu-Xin Han, Xing Huang, Bo-Han Li, Yue-Zhou Li, Hong L\"u, Jian-Xin Lu, Bo Ning, Yuan Sun, Hou-Wen Wu, Jun-Bao Wu, Yi Yan, Hai-Tang Yang and Wen-Li Yang for useful conversations.
This work is supported by National Natural Science Foundation of China(Grants No. 11305125, No. 12047502), the Basic Research Program of Natural Science of Shaanxi Province(Grants No. 2019JM-026), and the Double First-class University Construction Project of Northwest University.

\appendix
\setcounter{equation}{0}
\setcounter{subsection}{0}
\renewcommand{\theequation}{A.\arabic{equation}}
\renewcommand{\thesubsection}{A.\arabic{subsection}}

\section{Another construction of $SO(2,d)$ action}
Alternatively, we can construct the $SO(2,d)$ action by uplifting $\rd \mathcal L_{\rm Palatini}$ as in \cite{Witten:1988hc}. Ignoring the dimensions, the exterior derivative of the Palatini Lagrangian density is
\begin{eqnarray}\label{dPalatini}
\!\!\!\!&&\!\!\!\!\rd\left[\epsilon_{a_1\cdots a_{D}}\left(\Theta^{a_1a_2}+\tfrac{D-2}{D}\ell^{-2}e^{a_1}\wedge e^{a_2}\right)\wedge e^{a_3}\wedge \cdots\wedge e^{a_{D}}\right]
\cr\!\!\!\!&=&\!\!\!\!(D-2)\left[\epsilon_{a_1\cdots a_{D}}\left(\Theta^{a_1a_2}+\ell^{-2}e^{a_1}\wedge e^{a_2}\right)\wedge \rD e^{a_3}\wedge e^{a_4}\wedge \cdots\wedge e^{a_{D}}\right] \,.~~~~~~~~~~~~
\end{eqnarray}
The corresponding $SO(2,d)$ uplifting is given by
\begin{eqnarray}\label{D+1lift}
V_{D+1}\!\!\!\!&=&\!\!\!\!\epsilon_{\hat\alpha_1\cdots\hat\alpha_{D+1}}F^{\hat\alpha_1\hat\alpha_2}\wedge F^{\hat\alpha_3\hat\alpha_4}\wedge \rD Y^{\hat\alpha_5}\wedge\cdots\wedge\rD Y^{\hat\alpha_{D+1}} ~~~~~~~~~~~~
\end{eqnarray}
which comes back to (\ref{dPalatini}) under the Einstein gauge (\ref{Egauge}) .
We can also check that this uplifted  $D+1$ form (\ref{D+1lift}) is indeed closed
\begin{eqnarray}
\!\!\!\!&&\!\!\!\!\rd\Big(\epsilon_{\hat\alpha_1\cdots\hat\alpha_{D+1}}F^{\hat\alpha_1\hat\alpha_2}\wedge F^{\hat\alpha_3\hat\alpha_4}\wedge \rD Y^{\hat\alpha_5}\wedge\cdots\wedge\rD Y^{\hat\alpha_{D+1}}\Big)
\cr\!\!\!\!&=&\!\!\!\!(D-3)\epsilon_{\hat\alpha_1\cdots\hat\alpha_{D+1}}F^{\hat\alpha_1\hat\alpha_2}\wedge F^{\hat\alpha_3\hat\alpha_4}\wedge F^{\hat\alpha_5}{}_{\hat\beta} Y^{\hat\beta}
\wedge\rD Y^{\hat\alpha_6}\wedge\cdots\wedge\rD Y^{\hat\alpha_{D+1}}
\cr\!\!\!\!&=&\!\!\!\!\tfrac{(D-3)(D-4)}{6}\epsilon_{\hat\alpha_1\cdots\hat\alpha_{D+1}}F^{\hat\alpha_1\hat\alpha_2}\wedge F^{\hat\alpha_3\hat\alpha_4}\wedge F^{\hat\alpha_5\hat\alpha_6} \wedge Y_{\hat\beta}\rD Y^{\hat\beta}\wedge\rD Y^{\hat\alpha_7}\wedge\cdots\wedge\rD Y^{\hat\alpha_{D+1}}
\cr\!\!\!\!&=&\!\!\!\!0
\,.~~~~~~~~~~~~
\end{eqnarray}
Thus it is locally exact $V_{D+1}=\rd \mathcal L$ and the corresponding $D$-dimensional gauge invariant Lagrangian density $\mathcal L$ must exist.
In fact, by noticing that
\begin{eqnarray}
\!\!\!\!&&\!\!\!\!\rd\Big(\epsilon_{\hat\alpha_1\cdots\hat\alpha_{D+1}}F^{\hat\alpha_1\hat\alpha_2}\wedge\cdots\wedge F^{\hat\alpha_{2m-1}\hat\alpha_{2m}}\wedge \rD Y^{\hat\alpha_{2m+1}}\wedge\cdots\wedge\rD Y^{\hat\alpha_{D}}Y^{\hat\alpha_{D+1}}\Big)
\cr\!\!\!\!&=&\!\!\!\!(-1)^D\epsilon_{\hat\alpha_1\cdots\hat\alpha_{D+1}}F^{\hat\alpha_1\hat\alpha_2}\wedge\cdots\wedge F^{\hat\alpha_{2m-1}\hat\alpha_{2m}}\wedge \rD Y^{\hat\alpha_{2m+1}}\wedge\cdots\wedge\rD Y^{\hat\alpha_{D+1}}
\cr\!\!\!\!&&\!\!\!\!+\tfrac{(-1)^{D}\left(\frac{D}2-m\right)}{m+1}\ell^2\epsilon_{\hat\alpha_1\cdots\hat\alpha_{D+1}}F^{\hat\alpha_1\hat\alpha_2}\wedge\cdots\wedge F^{\hat\alpha_{2m+1}\hat\alpha_{2m+2}} \wedge\rD Y^{\hat\alpha_{2m+3}}\wedge\cdots\wedge\rD Y^{\hat\alpha_{D+1}}
\,,~~~~~~~~~~~~
\end{eqnarray}
the explicit expression of $\mathcal L$ can be decided recursively. For even dimensions $D=2k$, we have
\begin{eqnarray}\label{ALageven}
\!\!\!\!&&\!\!\!\!\epsilon_{\hat\alpha_1\cdots\hat\alpha_{2k+1}}F^{\hat\alpha_1\hat\alpha_2}\wedge F^{\hat\alpha_3\hat\alpha_4}\wedge \rD Y^{\hat\alpha_5}\wedge\cdots\wedge\rD Y^{\hat\alpha_{2k+1}}
\cr\!\!\!\!&=&\!\!\!\!\rd\left[\sum_{m=2}^{k}\frac{(-1)^{D+m}\,2\ell^{2(m-2)}\Gamma\!\left(k-1\right)}{m!\,\Gamma\!\left(k-m+1\right)}\epsilon_{\hat\alpha_1\cdots\hat\alpha_{2k+1}}F^{\hat\alpha_1\hat\alpha_2}\wedge\cdots\wedge F^{\hat\alpha_{2m-1}\hat\alpha_{2m}}\wedge \rD Y^{\hat\alpha_{2m+1}}\wedge\cdots\wedge\rD Y^{\hat\alpha_{2k}}Y^{\hat\alpha_{2k+1}}\right]\,;~~~~~~~~
\end{eqnarray}
while for the odd dimensions $D=2k+1$, we have
\begin{eqnarray}\label{ALagodd}
\!\!\!\!&&\!\!\!\!\epsilon_{\hat\alpha_1\cdots\hat\alpha_{2k+2}}F^{\hat\alpha_1\hat\alpha_2}\wedge F^{\hat\alpha_3\hat\alpha_4}\wedge \rD Y^{\hat\alpha_5}\wedge\cdots\wedge\rD Y^{\hat\alpha_{2k+2}}
\cr\!\!\!\!&=&\!\!\!\!\rd\left[\sum_{m=2}^{k}\frac{(-1)^{D+m}\,2\ell^{2(m-2)}\Gamma\!\left(k-\frac12\right)}{m!\,\Gamma\!\left(k-m+\frac32\right)}
\epsilon_{\hat\alpha_1\cdots\hat\alpha_{2k+2}}F^{\hat\alpha_1\hat\alpha_2}\wedge\cdots\wedge F^{\hat\alpha_{2m-1}\hat\alpha_{2m}}\wedge \rD Y^{\hat\alpha_{2m+1}}\wedge\cdots\wedge\rD Y^{\hat\alpha_{2k+1}}Y^{\hat\alpha_{2k+2}}\right]
\cr\!\!\!\!&&\!\!\!\!+\frac{2\,(-1)^{D+k+1}\ell^{2(k-1)}\Gamma\!\left(k-\frac12\right)}{(k+1)!\,\Gamma\!\left(\frac12\right)}
\epsilon_{\hat\alpha_1\cdots\hat\alpha_{2k+2}}F^{\hat\alpha_1\hat\alpha_2}\wedge\cdots\wedge F^{\hat\alpha_{2k+1}\hat\alpha_{2k+2}}
\cr\!\!\!\!&=&\!\!\!\!\rd\Big[\sum_{m=2}^{k}\frac{(-1)^{D+m}\,2\ell^{2(m-2)}\Gamma\!\left(k-\frac12\right)}{m!\,\Gamma\!\left(k-m+\frac32\right)}
\epsilon_{\hat\alpha_1\cdots\hat\alpha_{2k+2}}F^{\hat\alpha_1\hat\alpha_2}\wedge\cdots\wedge F^{\hat\alpha_{2m-1}\hat\alpha_{2m}}\wedge \rD Y^{\hat\alpha_{2m+1}}\wedge\cdots\wedge\rD Y^{\hat\alpha_{2k+1}}Y^{\hat\alpha_{2k+2}}
\cr&&~~~~~~~~~+\frac{2\,(-1)^{D+k+1}\ell^{2(k-1)}\Gamma\!\left(k-\frac12\right)}{(k+1)!\,\Gamma\!\left(\frac12\right)}\Omega_{2k+1}\Big]
\,,~~~~~~~~
\end{eqnarray}
where
\begin{eqnarray}
\Omega_{2k+1}
\!\!\!\!&=&\!\!\!\!
\sum_{m=0}^k\frac{(k+1)!}{(k+m+1)\,m!\,(k-m)!}\epsilon_{\hat\alpha_1\cdots\hat\alpha_{2k+2}} A^{\hat\alpha_1\hat\alpha_2}\wedge
A^{\hat\alpha_3}{}_{\hat\beta_1}\wedge A^{\hat\beta_1\hat\alpha_4}\wedge\cdots\wedge A^{\hat\alpha_{2m+1}}{}_{\hat\beta_m}\wedge A^{\hat\beta_{m}\hat\alpha_{2m+2}}
\cr&&~~~~~~~~~~~~~~~~~~~~~~~~~~~~~~~~~~~~~~~~~~~~~~~~~~\wedge\rd A^{\hat\alpha_{2m+3}\hat\alpha_{2m+4}}\wedge\cdots\wedge \rd A^{\hat\alpha_{2k+1}\hat\alpha_{2k+2}}\,,~~~~~~~
\cr
\rd\Omega_{2k+1}
\!\!\!\!&=&\!\!\!\!\epsilon_{\hat\alpha_1\cdots\hat\alpha_{2k+2}}F^{\hat\alpha_1\hat\alpha_2}\wedge\cdots\wedge F^{\hat\alpha_{2k+1}\hat\alpha_{2k+2}}\,.~~~~~~~
\end{eqnarray}

On the other hand, we notice that the exterior derivative of the Lagrangian density in (\ref{GRGT}) is
\begin{eqnarray}
\!\!\!\!&&\!\!\!\!\rd\left\{\epsilon_{\hat\alpha_1\cdots \hat\alpha_{D+1}}\left[ F^{\hat\alpha_1\hat\alpha_2}-\tfrac{2}{D\ell^{2}} \rD Y^{\hat\alpha_1}\wedge  \rD Y^{\hat\alpha_2}\right]\wedge  \rD Y^{\hat\alpha_3}\wedge\cdots \wedge  \rD Y^{\hat\alpha_D}Y^{\hat\alpha_{D+1}}\right\}
\cr\!\!\!\!&=&\!\!\!\!(D-2)\epsilon_{\hat\alpha_1\cdots \hat\alpha_{D+1}}F^{\hat\alpha_1\hat\alpha_2}\wedge F^{\hat\alpha_3}{}_{\hat\beta} Y^{\hat\beta}\wedge\rD Y^{\hat\alpha_4}\wedge\cdots \wedge  \rD Y^{\hat\alpha_D} Y ^{\hat\alpha_{D+1}}
\cr\!\!\!\!&&\!\!\!\!+(-1)^{D}\epsilon_{\hat\alpha_1\cdots \hat\alpha_{D+1}}F^{\hat\alpha_1\hat\alpha_2}\wedge\rD Y^{\hat\alpha_3}\wedge\cdots \wedge  \rD Y^{\hat\alpha_D}\wedge\rD Y ^{\hat\alpha_{D+1}}
\cr\!\!\!\!&&\!\!\!\!-2\ell^{-2} \epsilon_{\hat\alpha_1\cdots \hat\alpha_{D+1}}F^{\hat\alpha_1}{}_{\hat\beta} Y^{\hat\beta}\wedge\rD Y^{\hat\alpha_2}\wedge\cdots \wedge  \rD Y^{\hat\alpha_D} Y ^{\hat\alpha_{D+1}}
\cr\!\!\!\!&=&\!\!\!\!\tfrac{(D-3)(D-2)}4\epsilon_{\hat\alpha_1\cdots \hat\alpha_{D+1}}F^{\hat\alpha_1\hat\alpha_2}\wedge F^{\hat\alpha_3\hat\alpha_4} Y_{\hat\beta}\wedge\rD Y^{\hat\beta}\wedge\rD Y^{\hat\alpha_5}\wedge\cdots \wedge  \rD Y^{\hat\alpha_D} Y ^{\hat\alpha_{D+1}}
\cr\!\!\!\!&&\!\!\!\!+\tfrac{D-2}4\epsilon_{\hat\alpha_1\cdots \hat\alpha_{D+1}}F^{\hat\alpha_1\hat\alpha_2}\wedge F^{\hat\alpha_3\hat\alpha_{D+1}} Y_{\hat\beta}\wedge\rD Y^{\hat\alpha_4}\wedge\cdots \wedge  \rD Y^{\hat\alpha_D} Y ^{\hat\beta}
\cr\!\!\!\!&&\!\!\!\!+(-1)^{D}\epsilon_{\hat\alpha_1\cdots \hat\alpha_{D+1}}F^{\hat\alpha_1\hat\alpha_2}\wedge\rD Y^{\hat\alpha_3}\wedge\cdots \wedge  \rD Y^{\hat\alpha_D}\wedge\rD Y ^{\hat\alpha_{D+1}}
\cr\!\!\!\!&&\!\!\!\!-(D-1)\ell^{-2}\epsilon_{\hat\alpha_1\cdots \hat\alpha_{D+1}}F^{\hat\alpha_1\hat\alpha_2} Y_{\hat\beta}\wedge\rD Y^{\hat\beta}\wedge  \rD Y^{\hat\alpha_3}\wedge\cdots \wedge  \rD Y^{\hat\alpha_D} Y ^{\hat\alpha_{D+1}}
\cr\!\!\!\!&&\!\!\!\!-\ell^{-2}\epsilon_{\hat\alpha_1\cdots \hat\alpha_{D+1}}F^{\hat\alpha_1\hat\alpha_{D+1}} Y_{\hat\beta}\wedge\rD Y^{\hat\alpha_2}\wedge\cdots \wedge  \rD Y^{\hat\alpha_D} Y^{\hat\beta}
\cr\!\!\!\!&=&\!\!\!\!\tfrac{(-1)^{D}(D-2)}4\ell^2\epsilon_{\hat\alpha_1\cdots \hat\alpha_{D+1}}F^{\hat\alpha_1\hat\alpha_2}\wedge F^{\hat\alpha_3\hat\alpha_4} \wedge\rD Y^{\hat\alpha_5}\wedge\cdots \wedge  \rD Y^{\hat\alpha_{D+1}}
\,.~~~~~
\end{eqnarray}
It gives rise to the same $V_{D+1}$ as in (\ref{D+1lift}).
Thus for $D>2$, the Lagrangian obtained in (\ref{ALageven}) and (\ref{ALagodd}) must be equivalent to the one in (\ref{GRGT}) up to total derivative terms. For $D=2$, the Lagrangian density in (\ref{GRGT}) itself is locally a total derivative term. 
In fact, the difference between the two types of Lagrangian is related to the global angular form \cite{Freed:1998tg}.

\end{document}